\documentclass[conference]{IEEEtran}
\usepackage{graphicx}
\usepackage{balance}
\usepackage{comment}
\usepackage[noend]{algorithmic}
\usepackage{algorithm}
\usepackage{amsmath}
\usepackage{subfig}

\ifCLASSINFOpdf
\else
\fi
\usepackage{array}
\begin{document}
%
\title{Fair Adaptive Data Rate Allocation and Power Control in LoRaWAN}

\author{\IEEEauthorblockN{Khaled Q. Abdelfadeel, Victor Cionca, Dirk Pesch}
\IEEEauthorblockA{Nimbus Centre, Cork Institute of Technology, Ireland\\
khaled.abdelfadeel@mycit.ie, victor.cionca@cit.ie, dirk.pesch@cit.ie}}


%


\maketitle

\begin{abstract}
In this paper, we present results of a study of the data rate fairness among nodes within a LoRaWAN cell. Since LoRa/LoRaWAN supports various data rates, we firstly derive the fairest ratios of deploying each data rate within a cell for a fair collision probability. LoRa/LoRaWAN, like other frequency modulation based radio interfaces, exhibits the \textit{capture effect} in which only the stronger signal of colliding signals will be extracted. This leads to unfairness, where far nodes or nodes experiencing higher attenuation are less likely to see their packets received correctly. Therefore, we secondly develop a transmission power control algorithm to balance the received signal powers from all nodes regardless of their distances from the gateway for a fair data extraction. Simulations show that our approach achieves higher fairness in data rate than the state-of-art in almost all network configurations.
\end{abstract}


%
\IEEEpeerreviewmaketitle

\section{Introduction}

LoRa/LoRaWAN is considered one of the Low Power Wide Area Networks \cite{raza2017low} that promise to connect massive numbers of low-cost wireless devices/nodes\footnote{We use device and node interchangeably.}, thousands per cell, in a simple star topology. Nodes operate with low energy consumption and data can be transmitted over long distances, e.g. many kilometers. 
The wide coverage area of LoRaWAN\footnote{From here on we use LoRaWAN to refer to the whole stack and network architecture that uses LoRa modulation as defined by the LoRa Alliance.} is due to its unique modulation, Long Range (\textit{LoRa}) modulation (subsection~\ref{lora}), which has a large link budget. 
LoRa provides multiple transmission parameters: Spreading Factor $SF$, Bandwidth $BW$, Coding Rate $CR$ and Transmission Power $TP$ that can be tuned to trade data rate for range, power consumption, or sensitivity.

Spreading codes associated with $SFs$ in LoRa are pseudo-orthogonal, thus LoRa can support simultaneous transmissions using different $SFs$ as long as none is received with significantly higher power than the others \cite{mikhaylov2017scalability} \cite{goursaud2015dedicated}, as otherwise the strongest signal suppresses weaker signals. Also, when multiple simultaneously transmitted signals have the same $SF$, the strongest signal will suppress the weaker signals if the power difference is sufficiently high \cite{bor2016lora}. This is known as the capture effect.

In our earlier work in \cite{abdelfadeel2018fadr} we showed that the capture effect and especially the imperfect-orthogonality of $SFs$ can make LoRaWAN an unfair system because of the near-far problem. Transmissions from nodes that are far from the gateway are not received when colliding with transmissions from nodes closer to the gateway that have significantly higher received power. This effect is magnified by LoRaWAN's large link budget leading to large power difference between transmissions from far and near nodes. Therefore, controlling the received signal power of all nodes is important to achieve fairness.

Another source of unfairness is the data rate assigned to a node. Each data rate, defined through the combination of $SF$, $BW$ and $CR$, experiences different airtime, thus different collision probability. The collision probability is higher when using slow data rate combinations and low when using fast combinations. Following these considerations, we propose a data rate allocation and $TP$ control algorithm, called FADR, to achieve a fair data rate for all nodes within a LoRaWAN cell while at the same time being energy efficient.

The contributions of this paper are as follows: we firstly formulate the general fairest data rate distribution to achieve a fair collision probability among all deployed data rates in a LoRaWAN cell. Then based on this distribution, we propose FADR, a data rate allocation and $TP$ control algorithm, to achieve a fair data rate independent of distance from the gateway while avoiding excessively high $TPs$ in order to reduce energy consumption. We provide detailed results, comparisons and discussions to show and explain how FADR performs under various network configurations. Overall, simulations show that FADR outperforms the state-of-art in almost all network configurations.

The remainder of this paper is organized as follows: Section~\ref{literaturereview} provides an overview of LoRa/LoRaWAN and highlights related work. Section~\ref{proposedapproach} describes FADR in detail. We present a detailed evaluation and discussion of FADR and comparison to the state-of-the-art in Section~\ref{evaluationanddiscussion}. Finally, Section~\ref{conclusion} presents the conclusions.

\begin{figure*}
  \label{airtimeandenergy}
  \vspace{-2em}
  \centering
  \subfloat[Airtime]{
    \includegraphics[width=0.6\columnwidth]{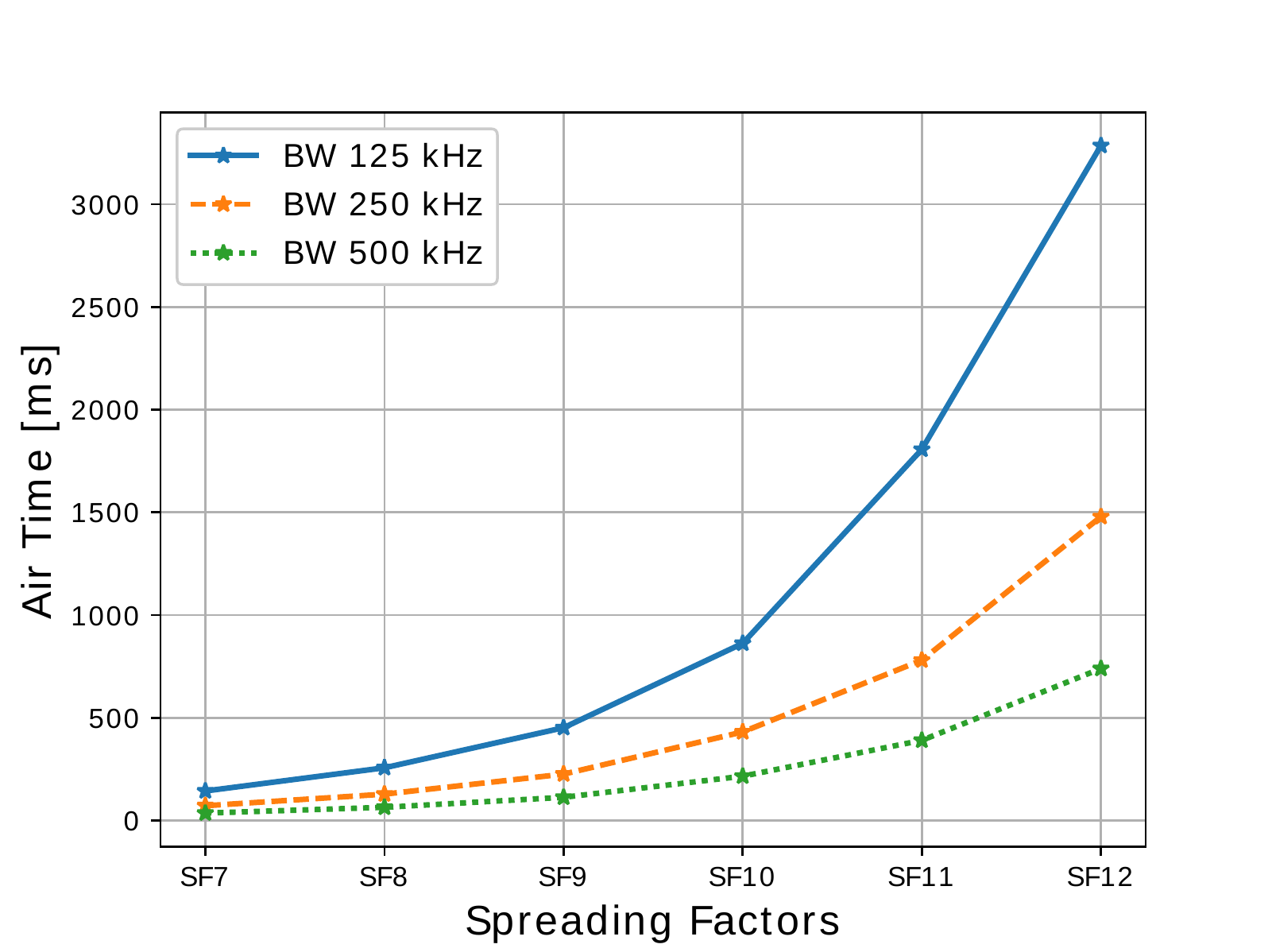}\label{fig:airtime_sf}
  }
  \subfloat[Energy]{
    \includegraphics[width=0.6\columnwidth]{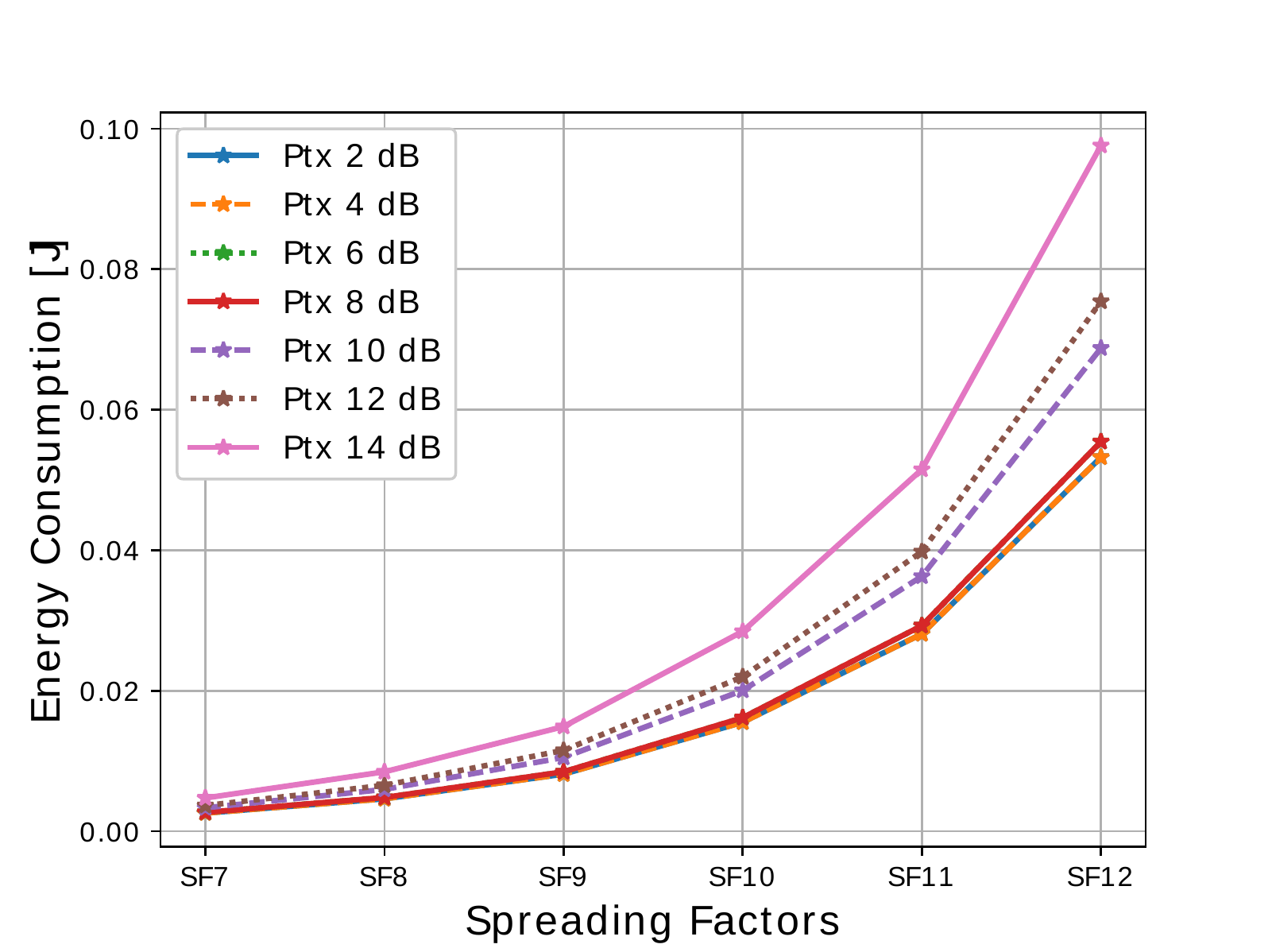}\label{fig:energy_sf}
  }
  \vspace{-0.5em}
  \caption{Effect of Spreading Factor on Airtime and Energy}
  \vspace{-0.5em}
\end{figure*}

\section{Literature Review} \label{literaturereview}
Here we provide an overview of the LoRaWAN protocol stack and highlight related work in the LoRaWAN domain.
\subsection{Long Range (LoRa)} \label{lora}

LoRa is a proprietary low-cost implementation of Chirp Spread Spectrum (CSS) modulation by Semtech that provides long range wireless communication with low power characteristics \cite{semtech2015lora} and represents the physical layer of the LoRaWAN stack. CSS uses wideband linear frequency modulated pulses, called \textit{chirps} to encode symbols. A LoRa symbol covers the entire bandwidth, making the modulation robust to channel noise and insensitive to frequency shifts.
LoRa modulation is defined by two main parameters: Spreading Factor $sf \in SFs (7,...,12)$, which affects the number of bits encoded per symbol, and Bandwidth $bw \in BWs (125,250,500) KHz$, which is the spectrum occupied by a symbol. A LoRa symbol consists of $2^{sf}$ chirps in which chirp rate equals bandwidth. 
LoRa supports forward error correction code rates $cr$ equal to $4/(4+n)$ where $n$ ranges from 1 to 4 to increase resilience. The theoretical bit rate $R_{b}$ of LoRa is shown in Eq.~\ref{eq1} \cite{semtech2015lora}.
\begin{equation}\label{eq1}
R_{b} = sf*\dfrac{bw}{2^{sf}}*cr \hspace{1cm}bits/s
\end{equation}
Moreover, a LoRa transceiver allows adjusting the Transmission Power $TP$. Due to hardware limitations the adjustment range is limited from $2dBm$ to $14dBm$ in $1dB$ steps.

A LoRa packet can be transmitted using a constant combination of $SF$, $BW$, $CR$ and $TP$, resulting in over 936 possible combinations. Tuning these parameters has a direct effect on the bit rate and hence the airtime, affecting reliability and energy consumption. Each increase in $SF$ nearly halves the bit rate and doubles the airtime and energy consumption but enhances the link reliability as it slows the transmission. Whereas each increase in the $BW$ doubles the bit rate and halves the airtime and energy consumption but reduces the link reliability as it adds more noise. 

The airtime of a LoRa packet can be precisely calculated by the LoRa airtime calculator \cite{lora2013calculator}. Fig.~\ref{fig:airtime_sf} shows the effect of $SFs$ and $BWs$ at code rate $CR=4/5$ on the airtime to transmit an 80 bytes packet length. As shown, the fastest combination uses the lowest $SF$ with the highest $BW$, whereas, the highest $SF$ with the lowest $BW$ achieves the slowest combination. Fig.~\ref{fig:energy_sf} shows the energy consumption for combinations of $SFs$ and $TPs$ at $CR=4/5$ and $BW=500KHz$ to transmit an 80 bytes packet. As shown, the $SF$ has much higher impact than the $TP$ on the energy consumption, e.g. increasing $SF$ consumes more energy than increasing $TP$ especially for large $SFs$.

LoRa modulation can enable concurrent transmissions, exploiting the pseudo-orthogonality of $SFs$ as long as none of the simultaneous transmissions is received with significantly higher power than the others \cite{goursaud2015dedicated}. Otherwise, the strongest transmission suppresses weaker transmissions if the power difference is higher than the Co-channel Interference Rejection (CIR) of weaker $SFs$. In case of the same $SF$, all simultaneous transmissions are lost, unless one of the transmissions is received with higher power than the CIR of the $SF$. This suppression of weaker signals by the strongest signal is called \textit{capture effect} \cite{bor2016lora}. The CIR of all $SF$ pairs has been calculated using simulations in \cite{goursaud2015dedicated} and validated by real LoRa link measurements in \cite{croce2017impact}.

\subsection{LoRaWAN} \label{lorawan}

LoRaWAN \cite{lorawan2017specs} is an open-source Medium Access Control (MAC) layer, system architecture and regional specifications using the LoRa modulation.
LoRaWAN MAC is based on simple Aloha, where a LoRa radio can transmit at any time as long as it respects the spectrum regulation. LoRaWAN operates in the Industrial Scientific and Medical (ISM) frequency band (868 MHz in Europe), which imposes a duty cycle of not more than 1\% on radios that do not adopt Listen-Before-Talk (LBT). The LoRaWAN system architecture is a simple star-of-stars topology where nodes communicate directly to one or more gateways which connect to a common network server. 
A LoRaWAN gateway is usually equipped with multiple LoRa transceivers, thus is able to receive multiple transmissions on all transmission parameter combinations at the same time. Therefore, a LoRa device can transmit data to a network server with any transmission parameter combination without any prior configuration.

\begin{table}
  \centering
  \caption{LoRaWAN Data Rates in Europe \cite{lorawan2017regionalparameters}} \label{tab1}
  \begin{tabular}{*{15}{c}}
  Data Rates & Parameter Combination & Indicative physical bit rate [bit/s]\\
  \hline
  0  & SF12 / 125 kHz & 250\\
  1  & SF11 / 125 kHz & 440\\
  2  & SF10 / 125 kHz & 980\\
  3  & SF9 / 125 kHz  & 1760\\
  4  & SF8 / 125 kHz  & 3125\\
  5  & SF7 / 125 kHz  & 5470\\
  6  & SF7 / 250 kHz  & 11000\\
  \hline
  \end{tabular}
\end{table}

LoRaWAN defines an Adaptive Data Rate (ADR) scheme to control the uplink transmission parameters of LoRa devices. A LoRa device expresses an interest in using the ADR scheme by setting the \textit{ADR} flag in any uplink MAC header. When the ADR scheme is enabled, the network server can control transmission parameters of a LoRa device using \textit{LinkADRReq} MAC commands. Typically, the network server collects the 20 most recent transmissions from a node, including Signal-to-Noise Ratio (SNR) and the number of gateways that received each transmission. Based on that history, the network server assigns transmission parameters to be more airtime and energy efficient. To reduce the \textit{LinkADRReq} command length, not all transmission parameters are available, but a subset of only 7 ($SF$ and $BW$) combinations as shown in table~\ref{tab1} and 5 $TPs$ (2,5,8,11, or 14) can be set \cite{lorawan2017specs}.

\subsection{Related Work}

Recent research on LoRa/LoRaWAN has mainly focused on LoRa performance evaluation in terms of coverage, capacity, scalability and lifetime. The studies have been carried out using real deployments in \cite{oliveira2017longrange} and \cite{juha2017performance}, mathematical models in \cite{bankov2017mathematical} and \cite{georgiou2017lpwan}, or computer simulations in \cite{bor2016lora} and \cite{margin2017performance}. Almost all these works have assumed perfectly orthogonal $SFs$ although it has been shown in \cite{mikhaylov2017scalability} and \cite{croce2017impact} that this is not a valid assumption.

Furthermore, recent work has proposed transmission parameter allocation approaches for LoRaWAN with different objectives. For example, authors in \cite{bor2017transmissionparameter} proposed a transmission parameter selection approach for LoRa to achieve low energy consumption at a specific link reliability. Here a LoRa node probes a link using a transmission parameter combination to determine the link reliability. It then chooses the next probe combination based on whether the new combination achieves lower energy consumption while maintaining at least the same link reliability. Finally, the approach terminates when reaching the optimal combination from an energy consumption perspective. 

Authors in \cite{cuomo2017explora} proposed two $SF$ allocation approaches, namely EXP-SF and EXP-AT, to help LoRaWAN achieve a high overall data rate. EXP-SF equally allocates $SFs$ to $N$ nodes based on the Received Signal Strength Indicator (RSSI), where the first $N/6$ nodes with the highest RSSI get $SF7$ assigned and then the next $N/6$ nodes $SF8$ and so on. EXP-AT is more dynamic than EXP-SF, where the $SF$ allocation theoretically equalizes the airtime of nodes. The two aforementioned works \cite{bor2017transmissionparameter} and \cite{cuomo2017explora} assumed perfectly orthogonal $SFs$, which leads to a higher overall data rate than in reality. 

In the context of our work presented here, allocating data rates and $TPs$ to achieve data rate fairness in LoRaWAN is not well investigated, with the exception of \cite{reynders2017power}, where authors proposed a power and spreading factor control approach to achieve fairness within a LoRaWAN cell. We provide an overview of \cite{reynders2017power} and a detailed comparison with our proposal in Section~\ref{evaluationanddiscussion}. While in general data rate and power control approaches have been well studied for cellular systems and WiFi \cite{yates1995uplinkpower} \cite{subramanian2005joint}, we argue that these solutions are not suitable for constrained systems like LoRaWAN. The reason is that cellular based approaches require fast feedback and high data rates to work, which are not available in LoRaWAN.

In the end, an interesting work was done to ensure an interoperability between LoRaWAN and the native IoT stack i.e. IPv6/UDP/CoAP at the device level. The interoperability was done by adopting legacy solution like 6LoWPAN over LoRaWAN \cite{weber2017ipv6} or by developing a new header compression technique to be more suitable for the constraints of LoRaWAN \cite{abdelfadeel2017lschc}.

\begin{figure*}
  \vspace{-2em}
  \centering
  \subfloat[Fairness Index]{
    \includegraphics[width=0.6\columnwidth]{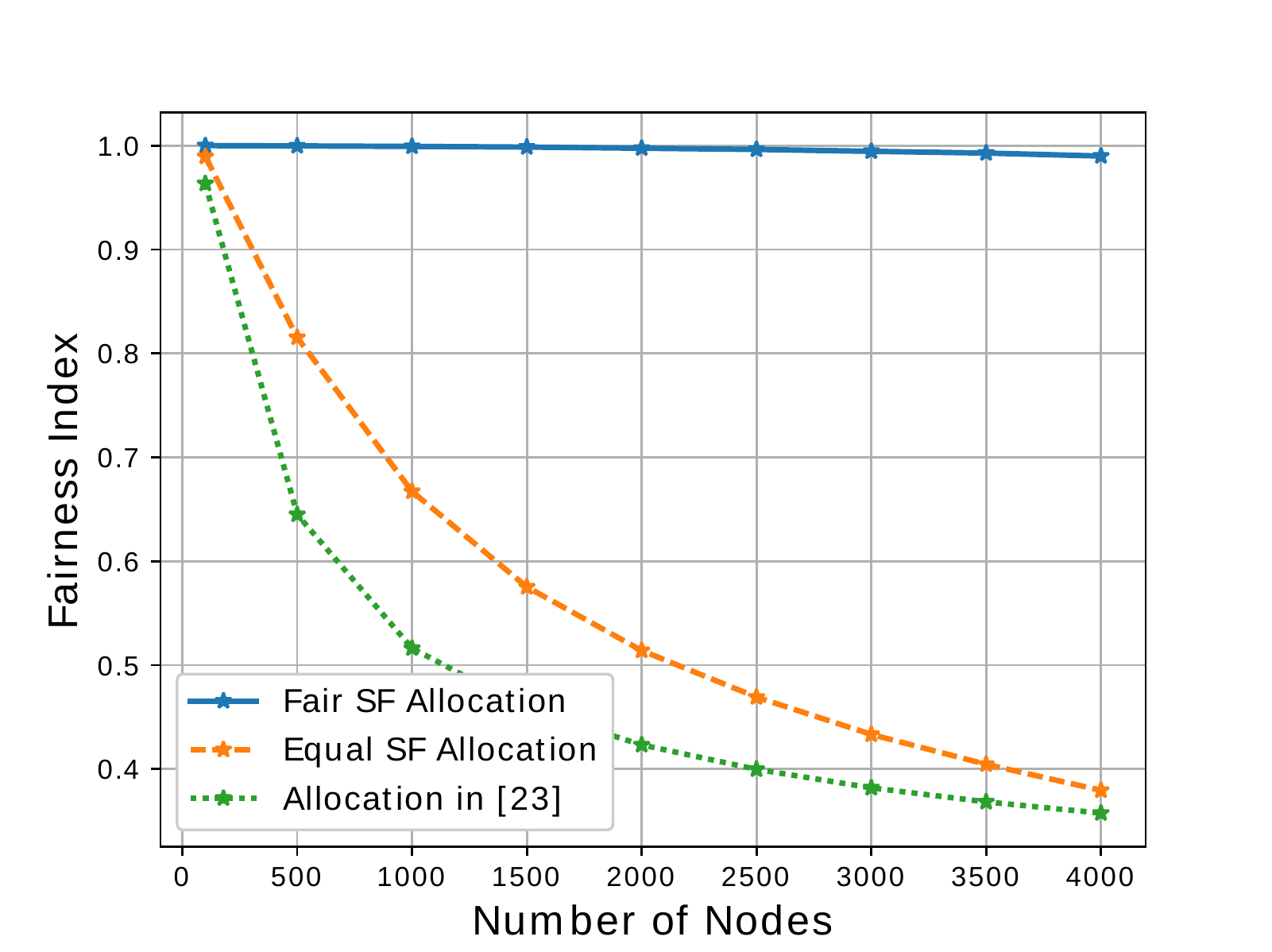}\label{fig:sfa_fairness}
  }
  \subfloat[DER]{
    \includegraphics[width=0.6\columnwidth]{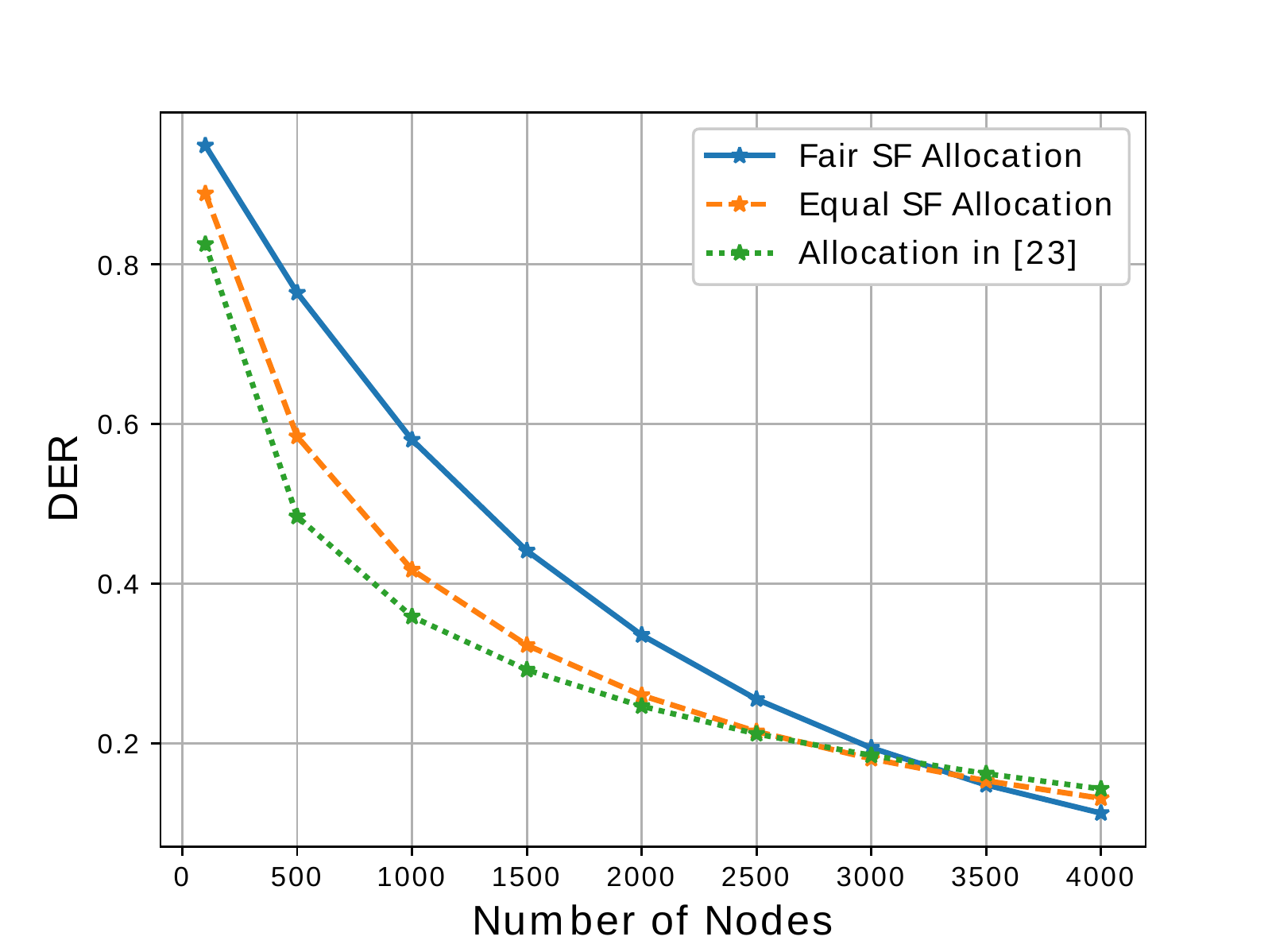}\label{fig:sfa_der}
  }
 \subfloat[DER vs SFs (4000 nodes)]{
    \includegraphics[width=0.6\columnwidth]{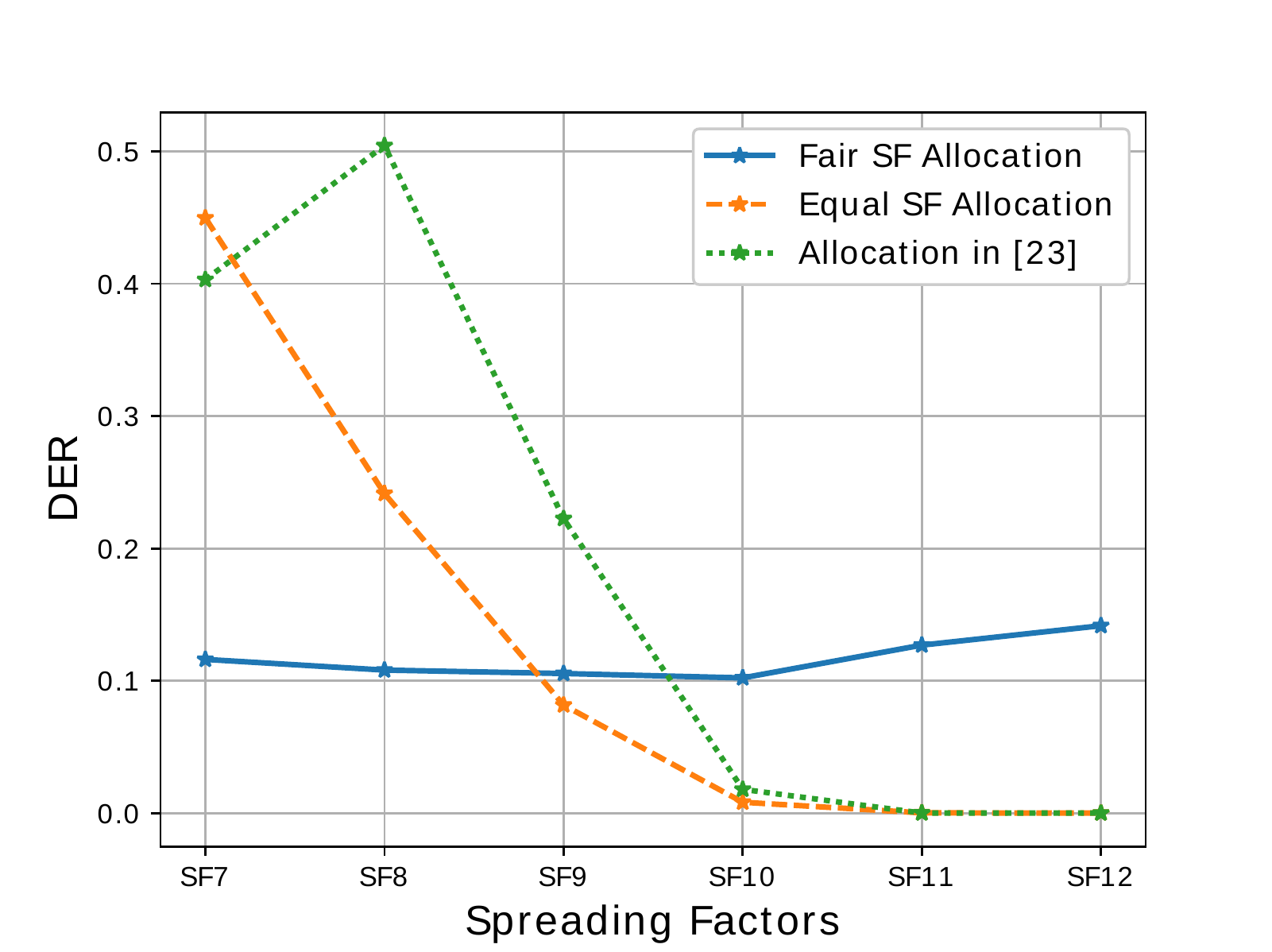}\label{fig:sfa_sfder}
  }
  \vspace{-0.5em}
  \caption{Different SF Allocations Study}\label{fig:sfastudy}
  \vspace{-0.5em}
\end{figure*}

\section{FADR Algorithm} \label{proposedapproach}
In the following we present our fair data rate allocation and power control proposal, which we call FADR, to achieve data rate fairness among nodes in a LoRaWAN cell. Firstly, we derive the fair data rate distribution in subsection~\ref{datarateallocation}, which tries to achieve an equal collision probability for all deployed data rates, then we provide our $TP$ control algorithm proposal in subsection~\ref{transmissionpowerallocation}, aimed at mitigating the capture and $SFs$ non-orthogonality effects.
\subsection{FADR - Data Rate Allocation} \label{datarateallocation}
Each transmission parameter combination ($SF$, $BW$, with $CR$) leads to a different data rate and thus airtime, which causes different collision probabilities, resulting in unfairness among nodes within a cell. Finding the fair data rate deployment ratios within a cell is therefore crucial.

$SF$ fair distribution ratios were derived in \cite{reynders2017power} as follows:
\begin{equation} \label{eq2}
p_{sf} = \frac{sf}{2^{sf}}/\sum_{i=7}^{12} \dfrac{i}{2^{i}} \hspace{1cm} \forall sf \in SFs,
\end{equation}
where $p_{sf}$ indicates the fraction of nodes using a specific $SF$. Eq.~\ref{eq2} has been derived by equalizing the collision probability of each $SF$ with taking into account the constraint that the sum of all probabilities must be unity $\sum_{s=7}^{12} p_{sf} = 1$.

However, Eq.~\ref{eq2} does not consider the impact of $BW$ and $CR$ on the collision probability. Assuming all $SFs$ will be deployed with the same $BW$ and $CR$ may not always be the case as the network operator may consider assigning different $BW$ and $CR$ to the same $SF$ in order to achieve a different data rate, reliability, or sensitivity. Therefore, we extend Eq.~\ref{eq2} into Eq.~\ref{eq3} to consider the impact of $BW$ in addition to $SF$ as follows:
\begin{equation} \label{eq3}
p_{sf,bw} = \dfrac{p_{sf}*bw}{\sum_{i \in BWs} i} \hspace{0.5cm} \forall sf \in SFs \hspace{0.1cm}\&\hspace{0.1cm} bw \in BWs,
\end{equation}
where $p_{sf,bw}$ indicates the fraction of nodes using a specific $SF$ and $BW$ combination. Eq.~\ref{eq3} is derived with respect to the constraint $\sum_{i \in BWs} p_{sf,bw} = p_{sf}$. In order to also consider the impact of $CR$, Eq.~\ref{eq3} is finally extended to Eq.~\ref{eq4} as follows:
\begin{equation} \label{eq4}
p_{sf,bw,cr} = \dfrac{p_{sf,bw}*cr}{\sum_{i \in CRs} i} \forall sf \in SFs \& bw \in BWs \& cr \in CRs,
\end{equation}
where $p_{sf,bw,cr}$ indicates the fraction of nodes using a specific $SF$, $BW$ and $CR$ combination. Eq.~\ref{eq4} is also derived with respect to the constraint $\sum_{i \in CRs} p_{sf,bw,cr} = p_{sf,bw}$. 

Eq.~\ref{eq4} is the generalized form of Eq.~\ref{eq2}, where in case of deploying all $SFs$ with the same $BW$ and $CR$, the values expressed by Eq. \ref{eq4} equal the values derived  with Eq.~\ref{eq2}. Hence, the fair ratios of using a potential LoRaWAN data rates as per table~\ref{tab1} without considering $CR$ are: $p_{0}=0.024$, $p_{1}=0.044$, $p_{2}=0.08$, $p_{3}=0.144$, $p_{4}=0.257$, $p_{5}=0.0898$, and $p_{6}=0.3592$.

Observing Eq.~\ref{eq4} in allocating the data rates within LoRaWAN cell ensures each node has the same probability of collision. However, this leaves the question as to what is the criteria of allocating data rates over RSSI within a cell? $BWs$ and $CRs$ are perfectly orthogonal, while the same is not true for $SFs$, which depend on the received power. We propose the region concept as a way of allocating $SFs$ within a LoRaWAN cell. For this a LoRaWAN cell is divided into regions, where each region consists of a number of nodes that are assigned to that region based on their RSSI. Nodes per region should be allocated using the fair $SF$ ratios from Eq.~\ref{eq2}. We recommend that the smallest size of a region should be equivalent to representing the smallest fair ratio, which for $SF12$ equals 2\% for a better representation of all ratios with in a region. Thus, we recommend the smallest region size should equal 50 nodes, which means $SF12$ is used by only one node in a region. We investigate the impact of region size in section~\ref{evaluationanddiscussion}.

To verify the impact of the fair data rate distribution, we compared the fair data rates, assuming all nodes are within a single region, versus equal $SF$ allocation across nodes, which has been considered in \cite{croce2017impact} and \cite{cuomo2017explora}, versus the proposed allocation in \cite{adelantado2017understanding}, where authors showed 28\% of nodes should use $SF12$. The fairness is calculated using Jain's fairness index \cite{jain1984quantitative}:
\begin{equation} \label{eq5}
\zeta = \frac{(\sum_{i=1}^N DER_i)^2}{N\sum_{i=1}^N DER_i^2},
\end{equation}
where $DER_i$ denotes the Data Extraction Rate (DER) of a node $i$ in a cell with $N$ nodes. The DER metric was introduced in \cite{bor2016lora} as the ratio of received packets to transmitted packets over a period of time. The fairness index varies from zero to one, where a higher index indicates a higher fairness. The results are shown in Fig.~\ref{fig:sfastudy} for different numbers of nodes, assuming perfectly orthogonal $SFs$ and neglecting the capture effect. This provides an insight into the fairness within a cell regardless of the assigned $TPs$.

Fig.~\ref{fig:sfa_fairness} shows the fairness index, where the fair allocation is almost one regardless of the number of nodes. However, it dramatically degrades in the other allocations with increasing number of nodes due to increasing collisions. The impact of allocation is clear in Fig.~\ref{fig:sfa_sfder}, which shows DER versus $SFs$ for a cell of 4000 nodes. The DER for nodes using a low $SF$ is higher than for those using a high $SF$ with equal $SF$ allocation and with $SF$ allocation as per \cite{adelantado2017understanding}. DER is almost zero for $SF10$, $SF11$ and $SF12$, which represent half of the nodes for the equal $SF$ allocation. This means, half of nodes cannot deliver any packets due to collisions, whereas the DER for the fair allocation is nearly equal for all $SFs$ and around the random access limit (see Fig.~\ref{fig:sfa_der}). The overall DER at the fair allocation outperforms the other two allocations up to about 3250 nodes. After that the overall number of collisions becomes higher, which means a lower DER than the other two allocations for the sake of equalizing the DER per $SF$ as shown in Fig.~\ref{fig:sfa_sfder}. 


\begin{algorithm}[!htp]
  \caption{FADR - $TP$ Control Algorithm}\label{alg:powercontrol}
  \begin{algorithmic}[1]
     \small
     \STATE   \textbf{Input} List of nodes \textbf{N}, corresponding \textbf{RSSI}, power levels \textbf{PowLevels}, matrix of \textbf{CIR}
     \STATE   \textbf{Output} $\forall n \in \textbf{N}, \textbf{P}[n] \in \textbf{PowLevels}$
     \STATE   Sort \textbf{N} by \textbf{RSSI}
     \STATE   \# Calculate MinRSSI, MaxRSSI, MinCIR
     \STATE   $MinRSSI = min(\textbf{RSSI})$, $MaxRSSI = max(\textbf{RSSI})$, $MinCIR=min(\textbf{CIR})$
     \STATE   $\textbf{PowLevels}.pop(0)$
     \FORALL  {$i \in \textbf{PowLevels}$}
     \STATE   $MaxPower = i$
     \IF      {$|MaxRSSI+MinPower-MinRSSI-MaxPower| <= MinCIR$}
     \STATE     $\textbf{PowLevels} = \textbf{PowLevels}[0:\textbf{PowLevels}.index(i)]$
     \STATE     $break$
     \ELSIF   {$i == max(\textbf{PowLevels})$}
     \STATE     $\textbf{powLevels}.pop()$
     \ENDIF
     \ENDFOR
     \STATE   \# Recalculate the minimum and the maximum of \textbf{RSSI}
     \STATE   $MinRSSI = min[MinRSSI + MaxPower, MaxRSSI + MinPower]$
     \STATE   $MaxRSSI = max[MinRSSI + MaxPower, MaxRSSI + MinPower]$
     \STATE   \# Assign the minimum power and save the MinPowIndex
     \FORALL  {$i \in range(0, len(\textbf{N}), 1)$}
     \IF          {$|\textbf{RSSI}[i]+MinPower| > |MinRSSI|$}
     \STATE            $MinPowIndex = i-1$ 
     \STATE            $break$
     \ELSE        
     \STATE       $\textbf{P}[i] = MinPower$
     \ENDIF
     \ENDFOR
     \STATE   \# Assign the maximum power and save the MaxPowIndex
     \FORALL  {$i \in range(len(N)-1, MinPowIndex,−1)$}
     \IF         {$|\textbf{RSSI}[i]+MaxPower-MinRSSI| > MinCIR$}
     \STATE            $MaxPowIndex = i-1$
     \STATE            $break$
     \ELSE       
     \STATE      $\textbf{P}[i] = MaxPower$
     \ENDIF
     \ENDFOR
     \STATE  \# Assign the nodes in between with the remaining power levels
     \STATE  $TempIndex = MinPowIndex$
     \FORALL {$i \in \textbf{PowLevels}$}
     \IF        {$(|\textbf{RSSI}[TempIndex]+i-MinRSSI| <= MinCIR)$\hspace{0.5cm}\AND$(|\textbf{RSSI}[TempIndex]+i-\textbf{RSSI}[MaxPowIndex]-MaxPower| <= MinCIR)$}
     \FORALL {$j \in range(TempIndex, MaxPowIndex, 1)$}
     \IF        {$|\textbf{RSSI}[j]+i-\textbf{RSSI}[MaxPowIndex]-MaxPower| > MinCIR$}
     \STATE              $TempIndex = j-1$ 
     \STATE              $break$
     \ELSE 
     \STATE     $\textbf{P}[j] = i$
     \ENDIF
     \ENDFOR     
     \ENDIF
     \ENDFOR
  \end{algorithmic}
\end{algorithm}

\subsection{FADR - Transmission Power Allocation} \label{transmissionpowerallocation}
The other aspect that creates unfairness in LoRaWAN is the near-far problem, which influences the capture effect, especially with not perfectly orthogonal $SFs$. These characteristics favor near nodes because of their higher received power than far nodes. Therefore, balancing the received powers of all nodes is required in order to achieve a fair data rate among all nodes regardless of their distance from the gateway.

Our proposed $TP$ control algorithm is shown in Algorithm~\ref{alg:powercontrol}. The algorithm requires a list of nodes (N) with a list of corresponding RSSIs, a list of available $TP$ levels (PowLevels) that can be assigned, and a matrix of CIR of all $SF$ pairs as inputs (line 1). To avoid RSSI instability, the algorithm is run after a certain number of packets have been collected by the network server in order to calculate the average RSSI. RSSI stability has been investigated in \cite{aref2014free}, which showed that the RSSI standard deviation of nodes close to the gateway is less than $3dBm$, however, the deviation increases to $20dBm$ for far nodes. Algorithm~\ref{alg:powercontrol} does not make assumptions on the initial $TP$ assignment of the collected packets, but recommends that nodes are initiated with the same $TP$ before running the algorithm to have RSSIs with a common reference.

Algorithm~\ref{alg:powercontrol} allocates a $TP$ to each node as output (line 2). The algorithm starts with sorting the nodes by their RSSI (line 3), next, calculates the maximum (MaxRSSI) and the minimum (MinRSSI) values of the measured RSSIs in addition to the minimum value of CIR (MinCIR), which represents the safe margin of all $SFs$ (line 5). Subsequently, the algorithm finds the maximum $TP$ (MaxPower) that can reduce the difference between RSSI extremes to below the safe margin (lines 6-13), where the minimum $TP$ (MinPower) is the minimum of PowLevels. In case that MaxPower is less than the maximum of PowLevels, the higher values are removed from the list because they will not be used (line 10). This will reduce the energy consumption, thus, extend the nodes' lifetime. Next, the algorithm assigns MinPower to the node with MaxRSSI and MaxPower to the node with MinRSSI, then recalculates MinRSSI and MaxRSSI accordingly (lines 15-16). Subsequently, the algorithm starts allocating the $TPs$ that can be divided into three stages. Firstly, allocating MinPower to high RSSI nodes as long as the new RSSI is not lower than the MinRSSI (lines 18-23). The index of the last node that complies with this approach is saved in MinPowIndex. Secondly, allocating MaxPower to low RSSI nodes as long as the new RSSI plus the safe margin is not higher than MinRSSI (line 25-30). The index of the last node that complies with this approach is saved in MaxPowIndex. Finally, the algorithm assigns to the nodes between MinPowIndex and MaxPowIndex the remaining $TPs$ from low to high as long as the allocation of each $TP$ complies with the rules that the new RSSI plus the safe margin is not lower than the first node using the same $TP$ (lines 33-40).

\begin{figure}
  \vspace{-2em}
  \centering   
  \includegraphics[width=0.9\columnwidth]{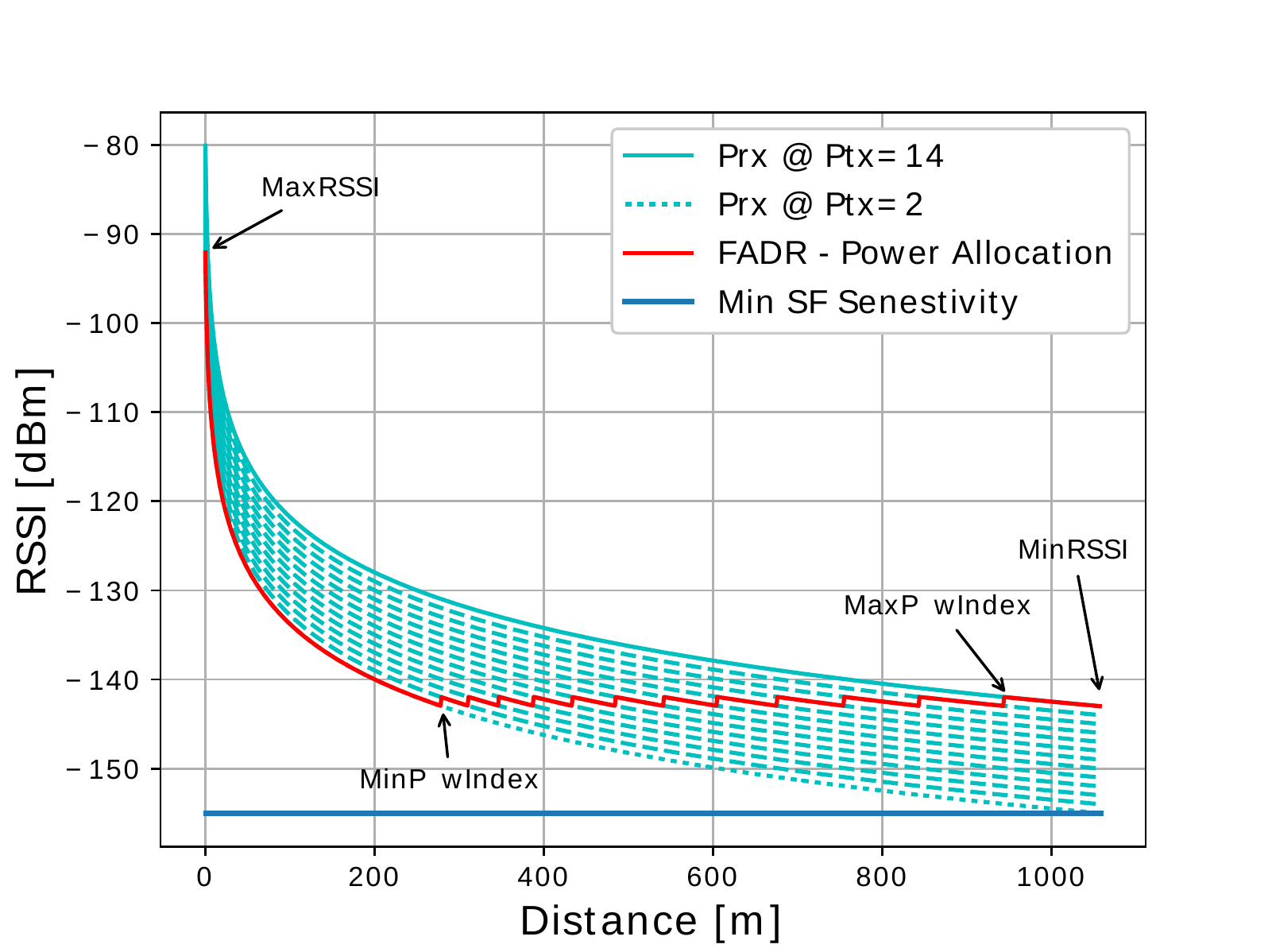}
  \vspace{-0.5em}
  \caption{FADR Power Allocation}\label{fig:fadr_powerallocation}
  \vspace{-0.5em}
\end{figure}

Fig.~\ref{fig:fadr_powerallocation} provides a visual example of how Algorithm~\ref{alg:powercontrol} works. In Fig.~\ref{fig:fadr_powerallocation}, MinPower is $2dBm$ and MaxPower is $14dBm$ (the maximum $TP$ for LoRaWAN) since the difference between RSSIs ($\sim50dBm$ in that example) is higher than the difference between $TPs$ ($12dBm$). Algorithm~\ref{alg:powercontrol} iterates forwardly to allocate MinPower until MinPowerIndex, then iterates backwardly to allocate MaxPower until MaxPowerIndex and finally iterates in between to assign the remaining $TPs$. We note that strong signal suppression of weaker signals effect can not be totally eliminated in all cases due to the limited, discrete $TP$ levels of LoRaWAN. However, Algorithm~\ref{alg:powercontrol} minimizes this effect as much as possible.

The run time of our algorithm is linear $O(N)$, where $N$ is the number of nodes per cell since the algorithm iterates over all nodes just once. This is important property because LoRaWAN potentially supports a massive number of nodes per cell. Therefore, our algorithm informally increases running time linearly with the nodes number.
\begin{figure*}
  \vspace{-2em}
  \centering
  \subfloat[Fairness Index]{
    \includegraphics[width=0.6\columnwidth]{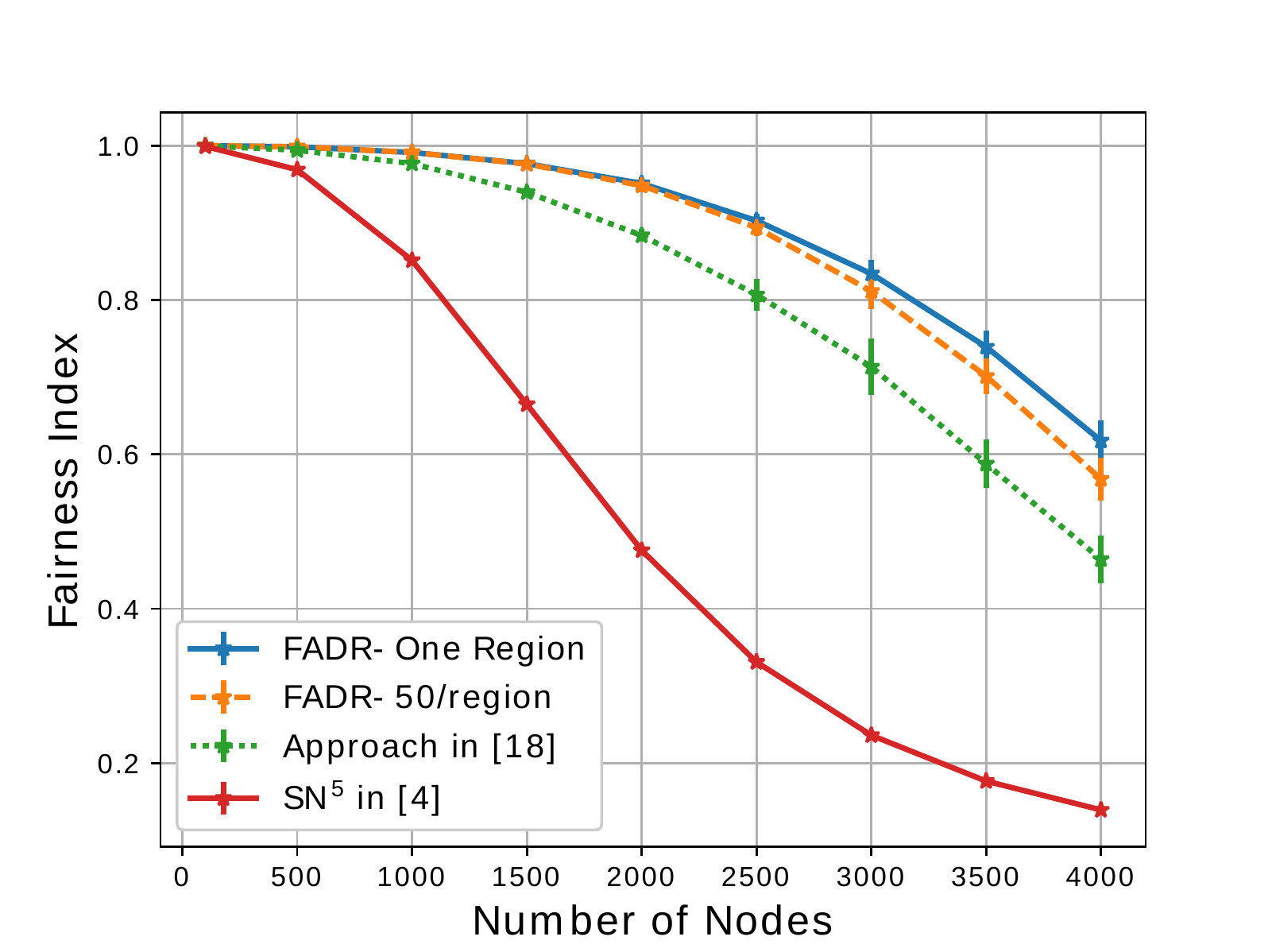}\label{fig:mc_fairness}
  }
  \subfloat[Overall DER]{
    \includegraphics[width=0.6\columnwidth]{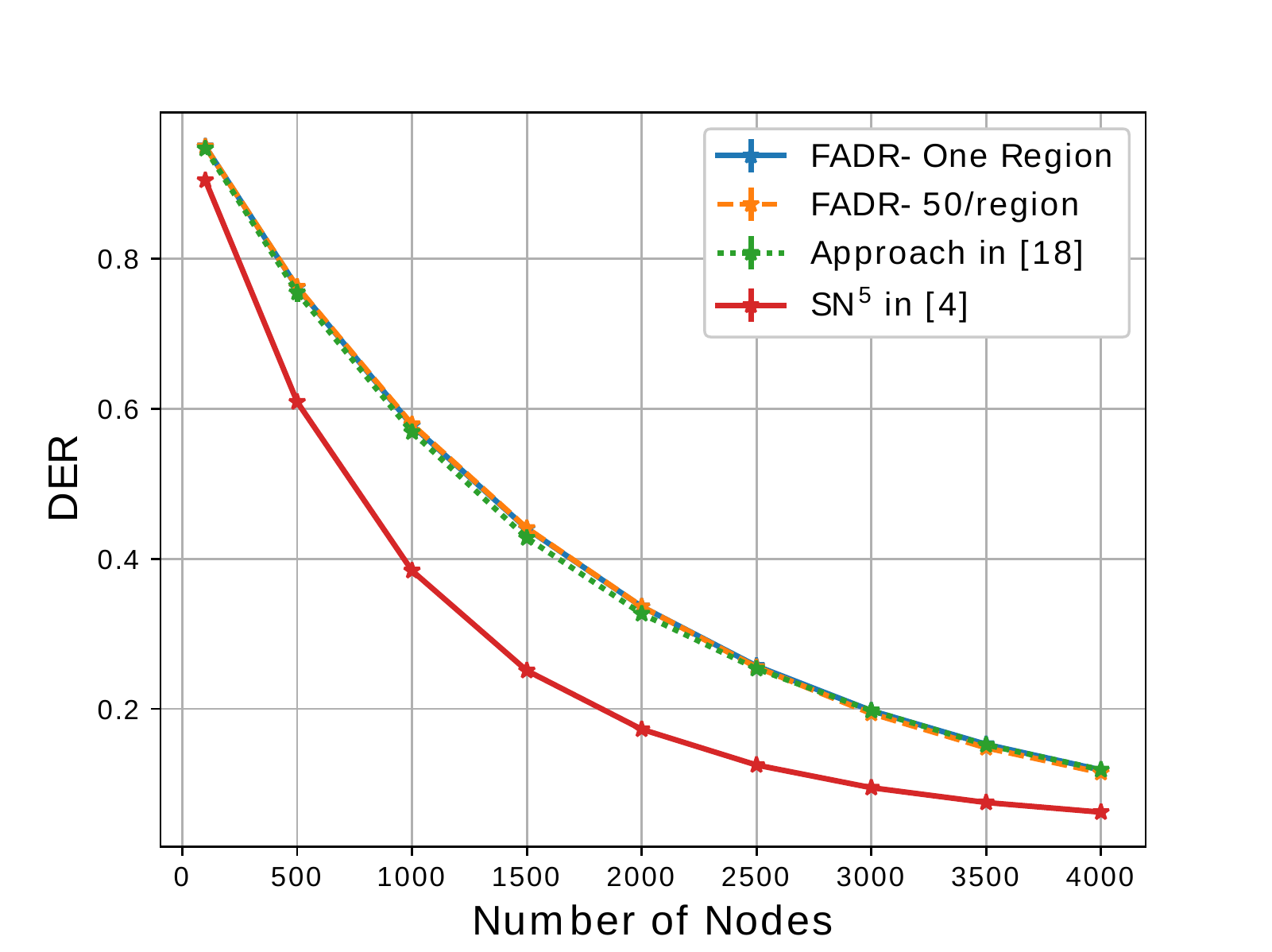}\label{fig:mc_der}
  }
 \subfloat[Overall Energy]{
    \includegraphics[width=0.6\columnwidth]{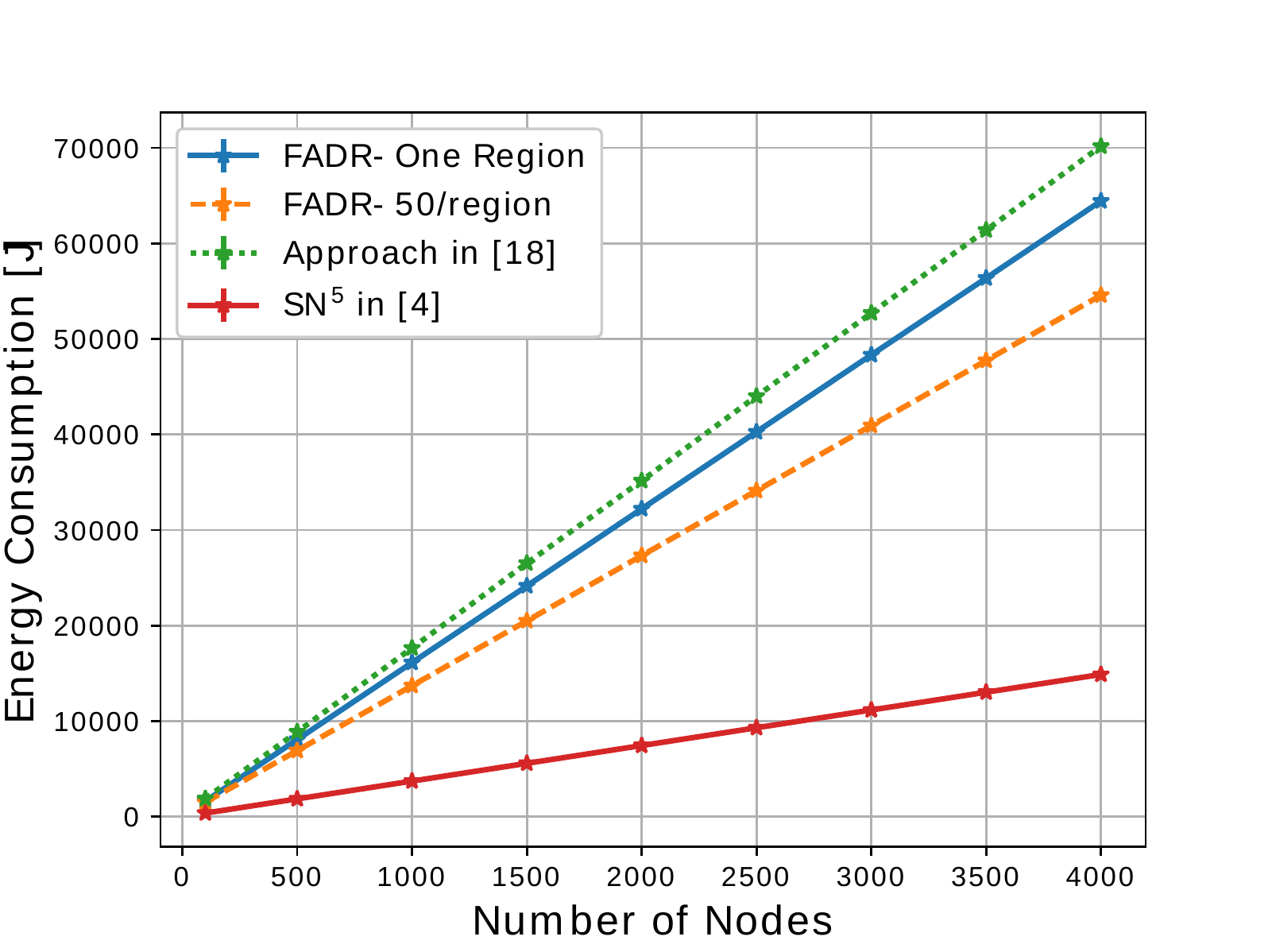}\label{fig:mc_energy}
  }
  \vspace{-0.5em}
  \caption{Main Comparison Results}\label{fig:maincomprsion}
  \vspace{-0.5em}
\end{figure*}

\section{Evaluation and Discussion} \label{evaluationanddiscussion}
To evaluate our ideas, we implemented FADR in LoRaSim \cite{bor2016lora}. LoRaSim is an open-source LoRa simulator that takes into account the capture effect only from the same $SF$, but otherwise assumes perfectly orthogonal $SFs$. In order to model collisions more comprehensively, we extended LoRaSim to include the non-perfect orthogonality property of $SFs$ based on the work in \cite{croce2017impact}, which adds a conservative $6dBm$ CIR threshold to all $SF$ pairs. We compared FADR to state-of-art approaches \cite{reynders2017power} and \cite{bor2016lora} by conducting multiple experiments that examine almost all factors that affect the algorithm. All experiments were run for a real-time of one day and repeated 10 times with different random seeds.

\subsection{State-of-the-art}
Authors of \cite{reynders2017power} present a $SF$ and $TP$ control algorithm to optimize the packet error rate fairness of LoRaWAN. The $SFs$ are allocated by sorting nodes first by their path loss, then by using the optimal distribution ratios Eq.~\ref{eq2}, where nodes with the lowest path loss get the lowest $SF$. The $TP$ control of \cite{reynders2017power} is based on the observation that nodes with high path loss and $SF8$ are the nodes with the highest packet error rate. Therefore, the algorithm assigns a high enough $TP$ to these nodes and allocates $SF7$ and $TP=2dBm$, i.e. short airtime and low $TP$, to all nodes that can corrupt these nodes' packets. Then the algorithm iterates again over all nodes to allocate enough $TP$ to all remaining nodes. We argue that this observation depends on the node distribution around the gateway, where these nodes may have lower or higher path loss depending on their locations from the gateway. We show when this assumption can be valid later on.

Authors of \cite{bor2016lora} show in their $SN^{5}$ experiment a way of allocating data rate and $TP$ in which each node chooses its transmission parameter combination locally to minimize first the airtime and secondly the lifetime. A node uses a combination that ensures its packets are received by the gateway and at the same time consumes less energy.

\subsection{Cell lay-out}
In this work, we consider a LoRaWAN cell that consists of one gateway located in the cell center and $N$ nodes placed randomly around the gateway. We investigated various cell radii $R$ and various number of nodes that are placed in the cell using different node distributions. Nodes generate data packets of length $L$ using transmission rate $\lambda$. A gateway is able to receive a configurable number of concurrent signals $MaxRecv$, based on its  number of LoRa transceivers, on the same carrier frequency $CF$ as long as concurrent transmissions use different $SFs$ and are within the safe margin. For a given combination of $SF$ and $BW$, packets are only decoded by the gateway if their RSSI is higher than the corresponding sensitivity. 

We used LoRaSim's propagation model which is based on the log-distance propagation model to calculate the RSSI of a node that transmits with $TP$. The same propagation model is used in \cite{croce2017impact} and \cite{margin2017performance}. Authors of \cite{croce2017impact} assume that any node, using any transmission parameter combination, is able to reach the gateway regardless of its distance from the gateway. To achieve the same assumption, the minimum sensitivity of all $SF$ and $BW$ combinations in LoRaSim was lowered to $-155dBm$, so that all nodes can reach the gateway with all combinations. Simulation parameters are shown in table~\ref{tab:simulationparameters}.

\begin{table}
  \centering
  \caption{Simulation Parameters} \label{tab:simulationparameters}
  \begin{tabular}{*{15}{c}}
  Parameter & Value & Unit\\
  \hline
  Nodes [$N$]  & 100-4000 &  \\
  Packet Length [$L$]  &  80  & byte \\
  Transmission Rate [$\lambda$]  & 60 & sec \\
  Max Reception [$MaxRecv$]  &  8  &   \\
  Cell Radius [$R$] & 100-3200 & m \\
  Channel Number &  1  &   \\
  Channel Frequency [$CP$] & 868 & MHz \\
  Simulation Time  &  86400 & sec \\
  Random Seeds &  10  &  \\
  \hline
  \end{tabular}
\end{table}


\subsection{Evaluation Experiments}
We conducted various experiments to show the performance evaluation of FADR versus the state-of-the-art. We firstly present the main performance evaluation in Sec.~\ref{maincomparison}. Then, the results are discussed in depth in Sec.~\ref{versusdistance}. Subsequently, the impact of the cell size is shown in Sec.~\ref{cellsize} and finally the impact of the node distribution is shown in Sec.~\ref{nodedistribution}.

\begin{figure*}
  \vspace{-2em}
  \centering
  \subfloat[DER vs Distance]{
    \includegraphics[width=0.6\columnwidth]{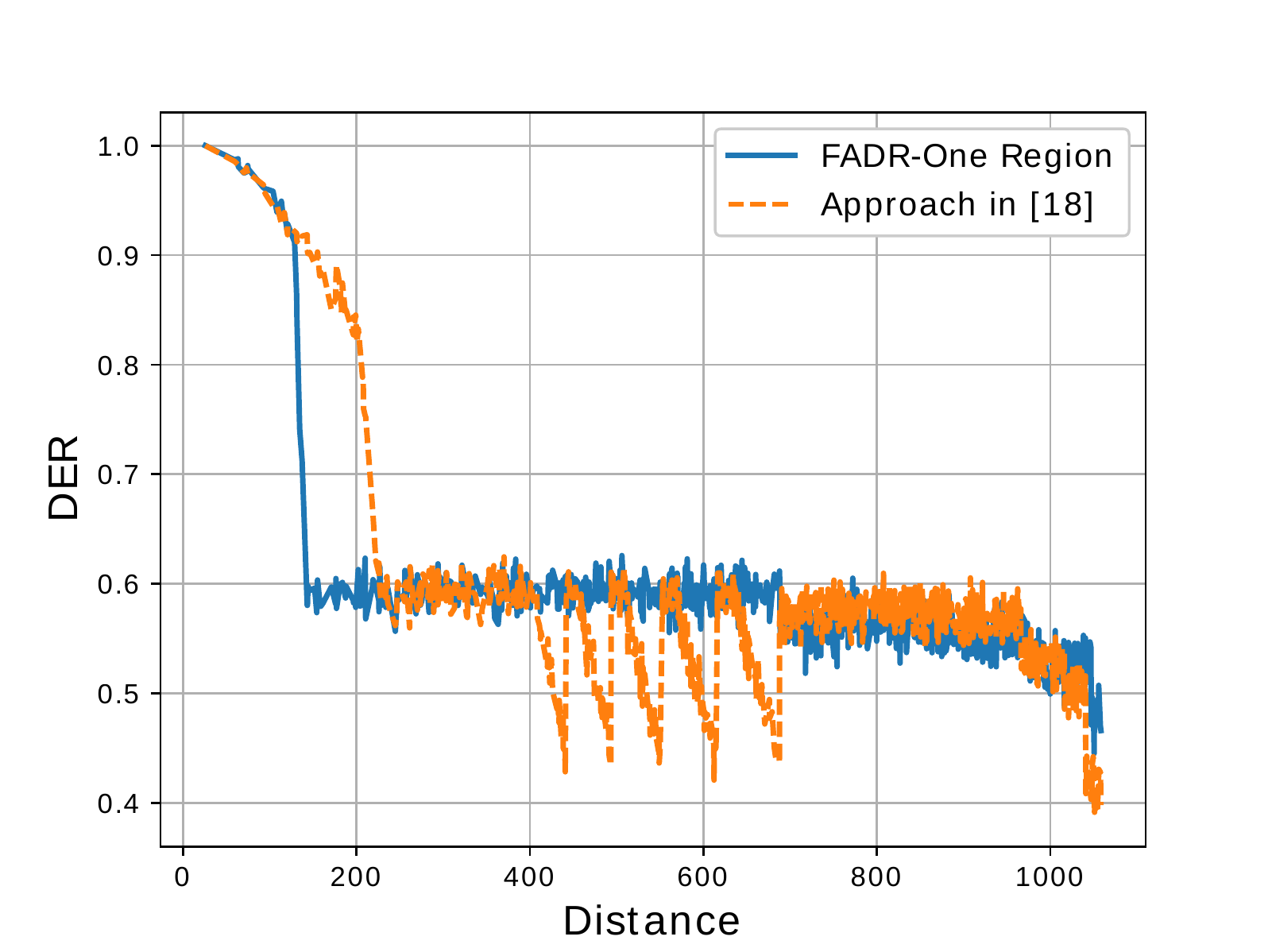}\label{fig:vd_der}
  }
 \subfloat[DER vs SFs]{
    \includegraphics[width=0.6\columnwidth]{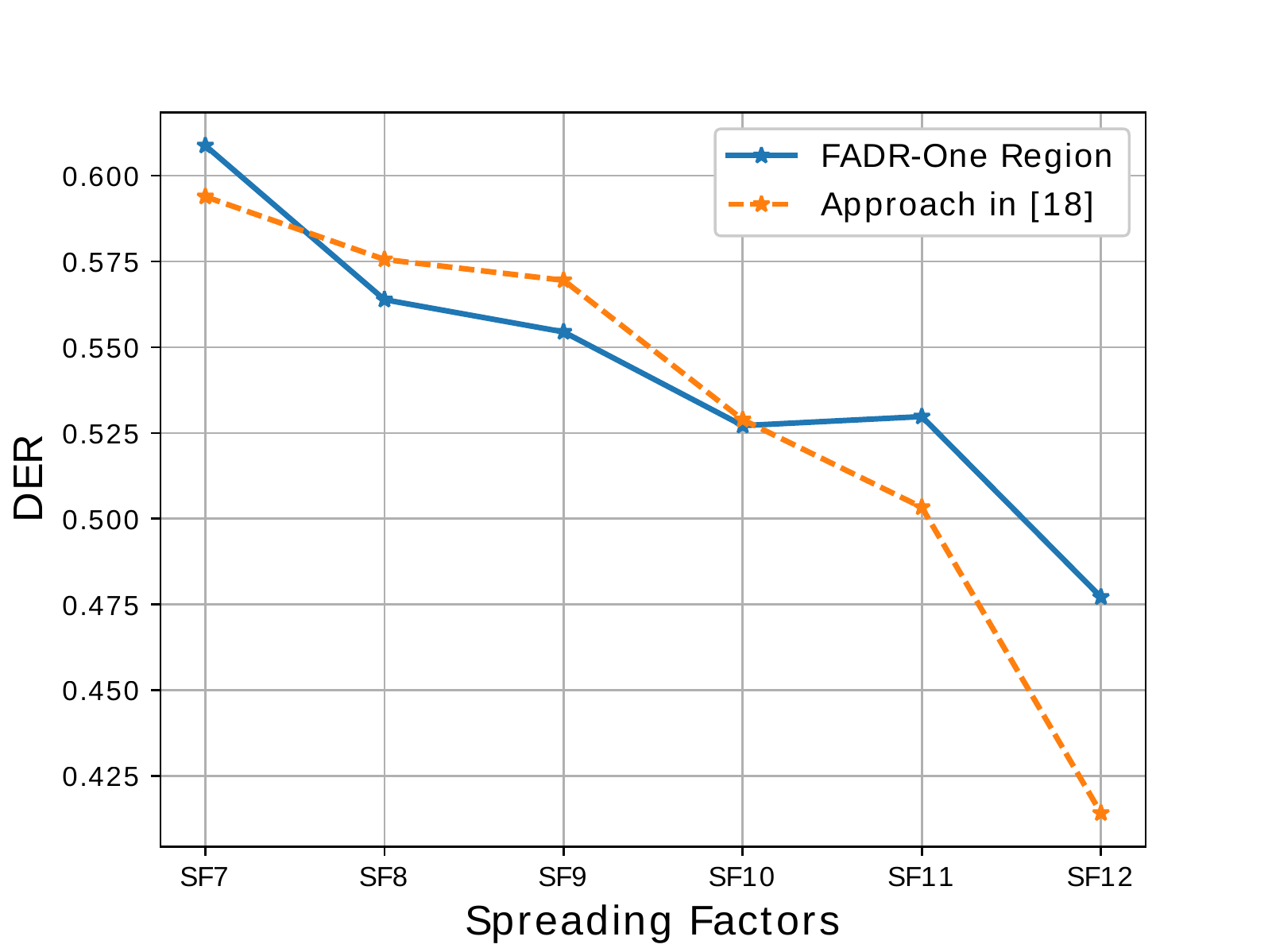}\label{fig:vd_sfder}
  }
 \subfloat[Transmission Power]{
    \includegraphics[width=0.6\columnwidth]{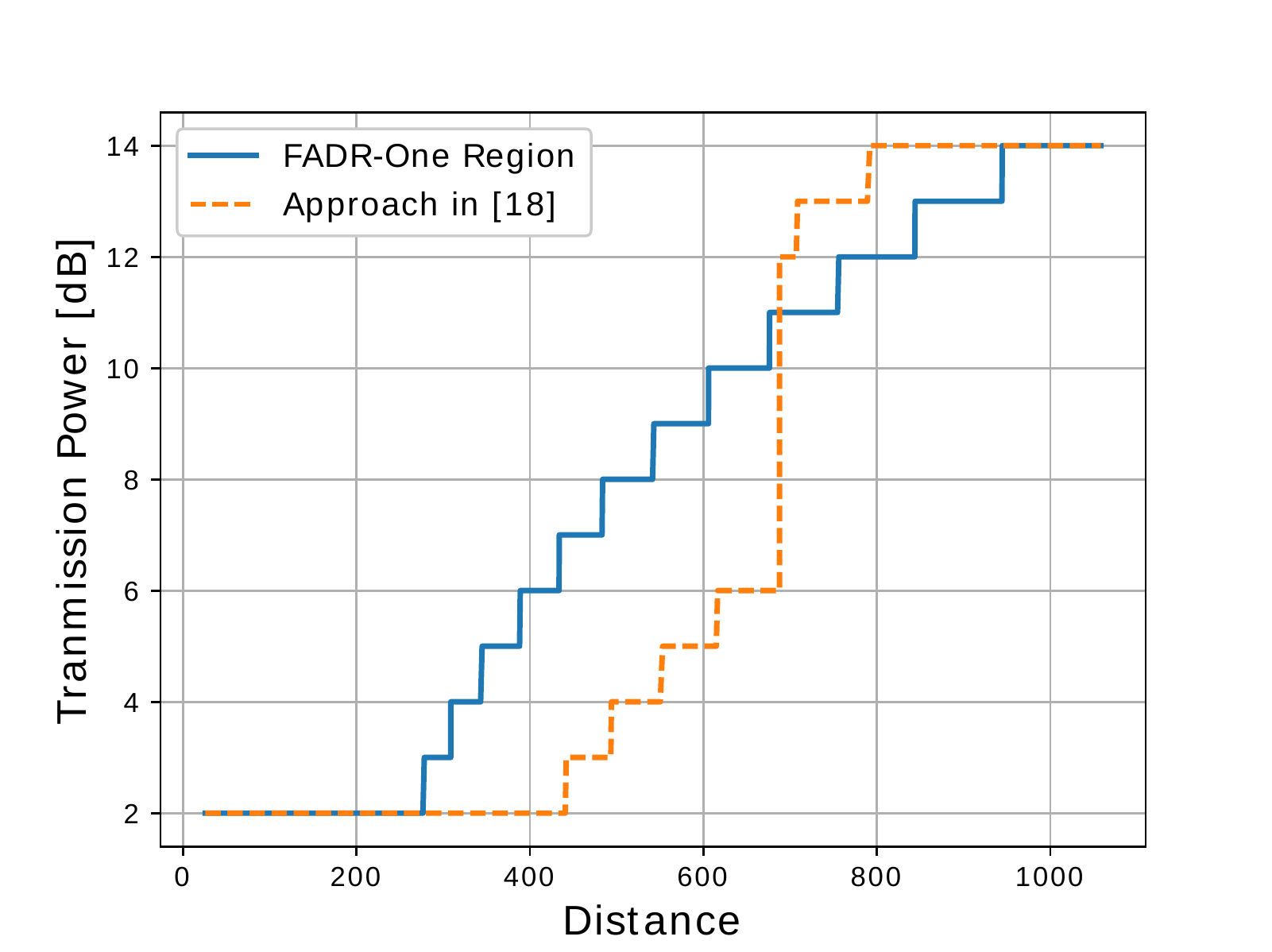}\label{fig:vd_txs}
  }
  \vspace{-1em}
  \\
 \subfloat[Spreading Factor]{
    \includegraphics[width=0.6\columnwidth]{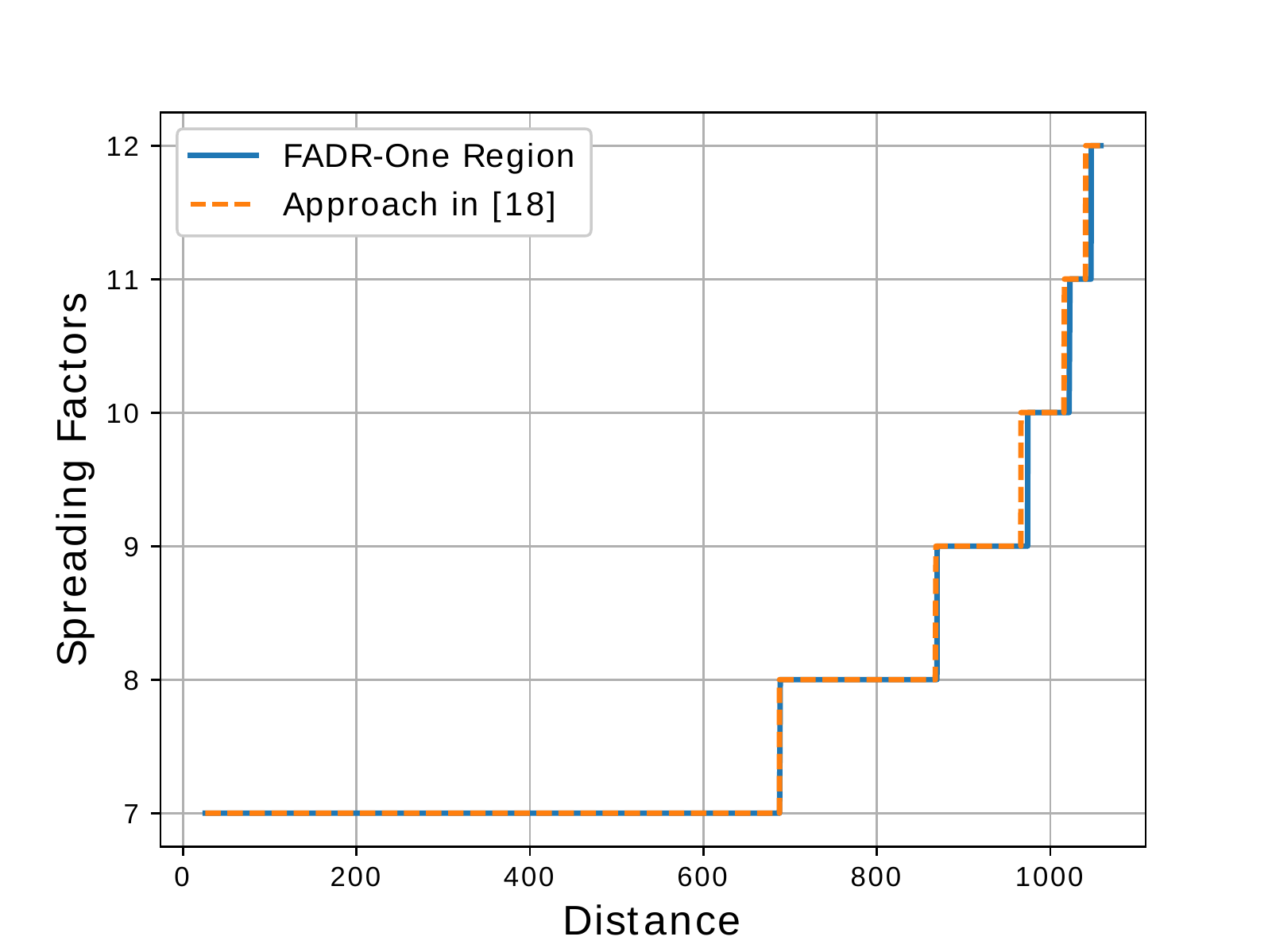}\label{fig:vd_sfdist}
  }
 \subfloat[RSSI vs Distance]{
    \includegraphics[width=0.6\columnwidth]{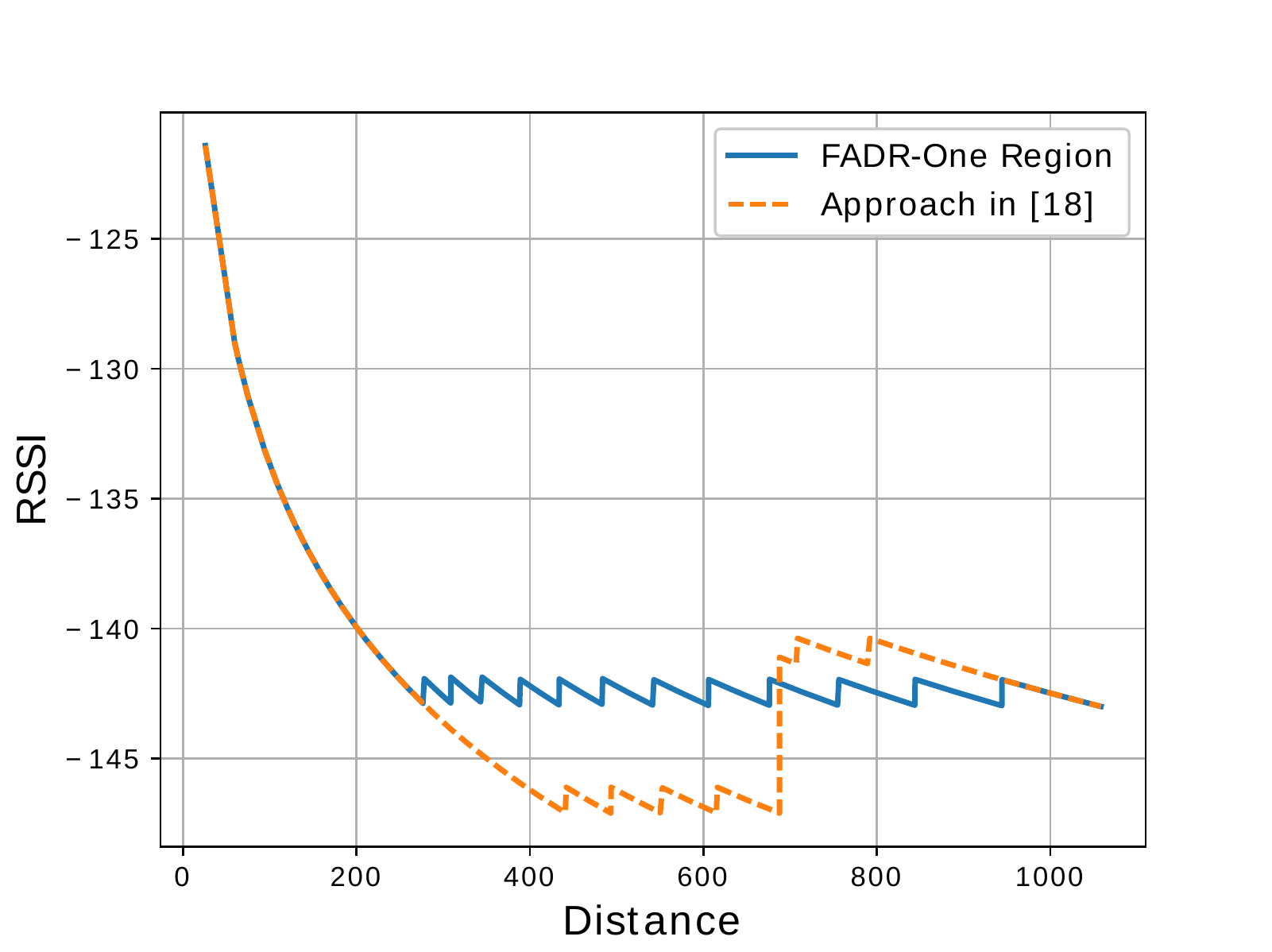}\label{fig:vd_rssi}
  }
  \vspace{-0.5em}
  \caption{Distance Study}\label{fig:vdstudy1}
  \vspace{-0.5em}
\end{figure*}

\subsubsection{Main Comparison}\label{maincomparison}
Fig.~\ref{fig:maincomprsion} shows the overall results of this study. Fig.~\ref{fig:mc_fairness} shows the fairness index using Eq.~\ref{eq5}, Fig~\ref{fig:mc_der} shows the overall DER, and Fig.~\ref{fig:mc_energy} shows the overall energy consumption. We evaluate FADR with two region configurations. 
First, in FADR-One Region, the entire cell is considered a single region. The second approach, where nodes are sorted according to their RSSI and then divided into groups of 50 nodes, was proposed in \cite{abdelfadeel2018fadr}. The data rate \textit{per region} is allocated based on Eq.~\ref{eq4} and $TP$ allocation is based on Algorithm~\ref{alg:powercontrol}.

Overall, both FADR region size approaches surpass the other approaches in terms of fairness without sacrificing the overall DER compared to \cite{reynders2017power} and with a remarkable improvement compared to $SN^{5}$ in \cite{bor2016lora}. On the other hand, both FADR region size approaches consume overall less energy than the approach in \cite{reynders2017power} but a higher energy than $SN^{5}$ in \cite{bor2016lora}, where all nodes choose to transmit using the lowest $TP$.

The low fairness and DER performance of $SN^{5}$ in \cite{bor2016lora} is due to the fact that data rate and $TP$ allocation was no studied at the cell level. Rather, nodes choose their transmission parameter locally, which leads to all nodes choosing the same transmission combination that achieves the lowest airtime and using the lowest $TP$, which achieves the lowest energy consumption, regardless of the cell status. This leads to a degradation of the cell performance by increasing the number of collisions within the cell and leads to aggressive unfairness for far nodes especially when increasing the number of nodes. In the following, we focus on the impact of the region size and analyze the performance of FADR versus \cite{reynders2017power} in the next subsection.

The region size has a notable impact on fairness and overall energy consumption, but almost no impact on the overall DER. Decreasing the region size, on one hand, mixes up all $SFs$ in a small variance of RSSI, on the other hand, $SFs$ are distributed everywhere in the cell, not just in contiguous areas as is the case in single region deployment, which allocates low $SFs$ to high RSSIs and high $SFs$ to low RSSIs. Therefore, small regions serve high $SFs$ better to the detriment of lower $SFs$, especially $SF7$, by decreasing the imperfect-orthogonality effect of $SF7$ over high $SFs$. However, small region deployment increases the impact of the capture effect, especially of $SF7$, because nodes with the same $SF$ now have high variance in their RSSIs. As overall nodes with $SF7$ represent the majority of nodes in a cell, small region deployment leads to lower fairness index, as shown in Fig~\ref{fig:mc_fairness}. However, excluding $SF7$ from the analysis and recalculating the fairness index shows that the small region deployment achieves higher fairness than single region deployment.

In terms of energy consumption, the FADR $TP$ control algorithm assigns high $TPs$ to low RSSI and vice versa, which leads to single region deployment consuming higher energy than small region deployments. The reason for this is that the nodes with low RSSI, in single region deployment, are allocated with high $SFs$, i.e. high airtime, and transmit using high $TPs$, but in small region deployments $SFs$ are distributed over the whole cell, thus, airtimes are distributed over $TPs$ as well.

\subsubsection{Distance Study}\label{versusdistance}

Figure ~\ref{fig:vdstudy1} shows DER (Fig.~\ref{fig:vd_der}), transmission powers (Fig.~\ref{fig:vd_txs}), $SF$ distribution (Fig.~\ref{fig:vd_sfdist}), and RSSIs (Fig.~\ref{fig:vd_rssi}) versus distance in addition to DER per $SF$ (Fig.~\ref{fig:vd_sfder}). These figures provide insights as to why FADR outperforms the approach published in \cite{reynders2017power}. The results of this study were collected from a cell with 1000 nodes, but we performed the same experiment with a larger number of nodes and got the same behavior.

FADR's advantage over \cite{reynders2017power} is shown in Fig.~\ref{fig:vd_der} in which FADR achieves roughly the same DER for a larger proportion of the network compared to \cite{reynders2017power} making FADR fairer. Between 400-700m, Reynders' approach \cite{reynders2017power} experiences high variation in the DER, corresponding to nodes using $SF7$ and low RSSIs. These nodes suffer from an aggressive capture effect by other nodes using $SF7$ and higher RSSIs and at the same time suffer from a capture effect due to the non-orthogonality of $SFs$ from nodes using different $SFs$ and higher RSSIs as shown in Fig.~\ref{fig:vd_rssi} because they do not get enough $TP$ as shown in Fig.~\ref{fig:vd_txs}.

It seems that the $TP$ control algorithm in \cite{reynders2017power} provides a $TP$ boost to nodes with $SF8-12$ over nodes with $SF7$ and low RSSIs as shown in Fig.~\ref{fig:vd_txs}. This boost yields an advantage to nodes with high $SFs$ over low RSSI nodes with $SF7$ (these low RSSI nodes suffer from low DER as shown in Fig.~\ref{fig:vd_der}) by reducing their non-orthogonality impact on high $SFs$. However, this boost creates a non-orthogonality impact from $SF8-9$ over higher $SFs$ if their RSSIs surpass RSSI of nodes using higher $SFs$ by the safe margin as shown in Fig.~\ref{fig:vd_sfder}. Because fewer nodes use $SF10-12$ than use $SF8-9$, \cite{reynders2017power} has slightly higher overall DER, but lower fairness than FADR. This $TP$ boost is the reason for a higher energy consumption compared to FADR.

On the other hand, our FADR $TP$ control algorithm increases the $TP$ gradually and within the safe margin after reaching the minimum limit of using the minimum $TP$ independently of the $SF$. This ensures that a large proportion of distances around the gateway have a balanced RSSI within the safe margin. With FADR, the nodes close to the gateway have an equal impact over the rest of the cell's nodes. This leads to a slight reduction in the overall DER. However, the DER will be more uniform over distance leading to higher fairness as shown in Fig.~\ref{fig:vd_der} and Fig.~\ref{fig:vd_sfder}.

\begin{figure}
  \vspace{-2em}
  \centering
  \includegraphics[width=0.6\columnwidth]{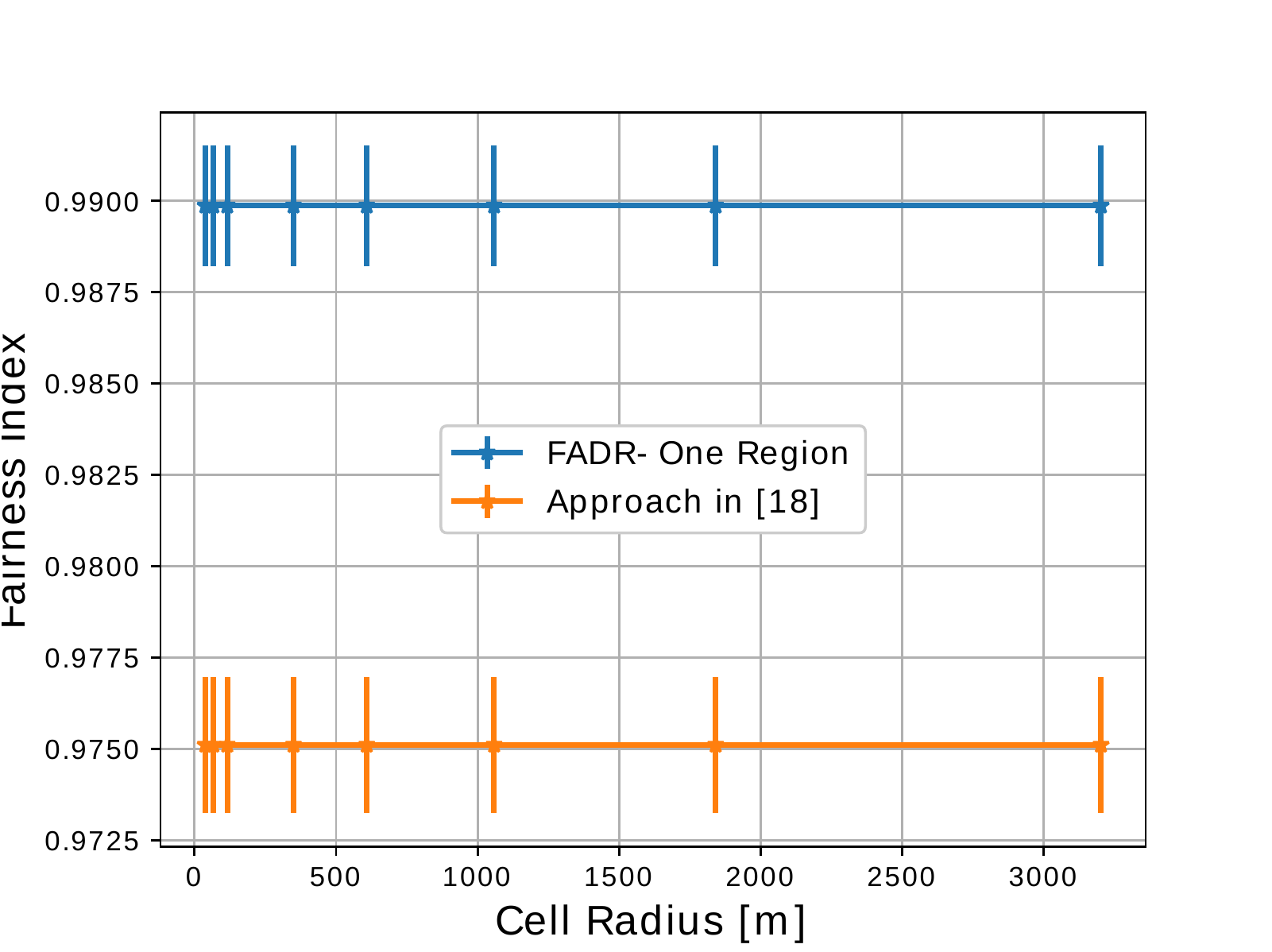}
  \vspace{-0.5em}
  \caption{Cell Size Study}\label{fig:cs_fairness}
  \vspace{-0.5em}
\end{figure}

\subsubsection{Cell Size Study}\label{cellsize}
We investigated the impact of the cell radius, while keeping the number of nodes constant, on the fairness, with results shown in Fig.~\ref{fig:cs_fairness}. Increasing the cell radius should provide an increase in the difference of nodes' RSSIs at the gateway. The result shown are collected from a cell with 1000 nodes. However, the behavior is identical to scenarios with other number of nodes. As shown, the cell radius does not have any impact on the fairness of either algorithms, where the difference is always the same. The reason for this is the slow increase in path loss when moving further away. For example, the difference of the RSSIs experienced at $1Km$ cell radius is ca. $62dBm$ and ca. $72dBm$ for $3Km$ cell radius. The $10dBm$ difference between the two cell radii can be handled well within the safe margin of either algorithms.

\begin{figure*}
  \vspace{-2em}
  \centering
  \subfloat[Overall DER]{
      \includegraphics[width=0.6\columnwidth]{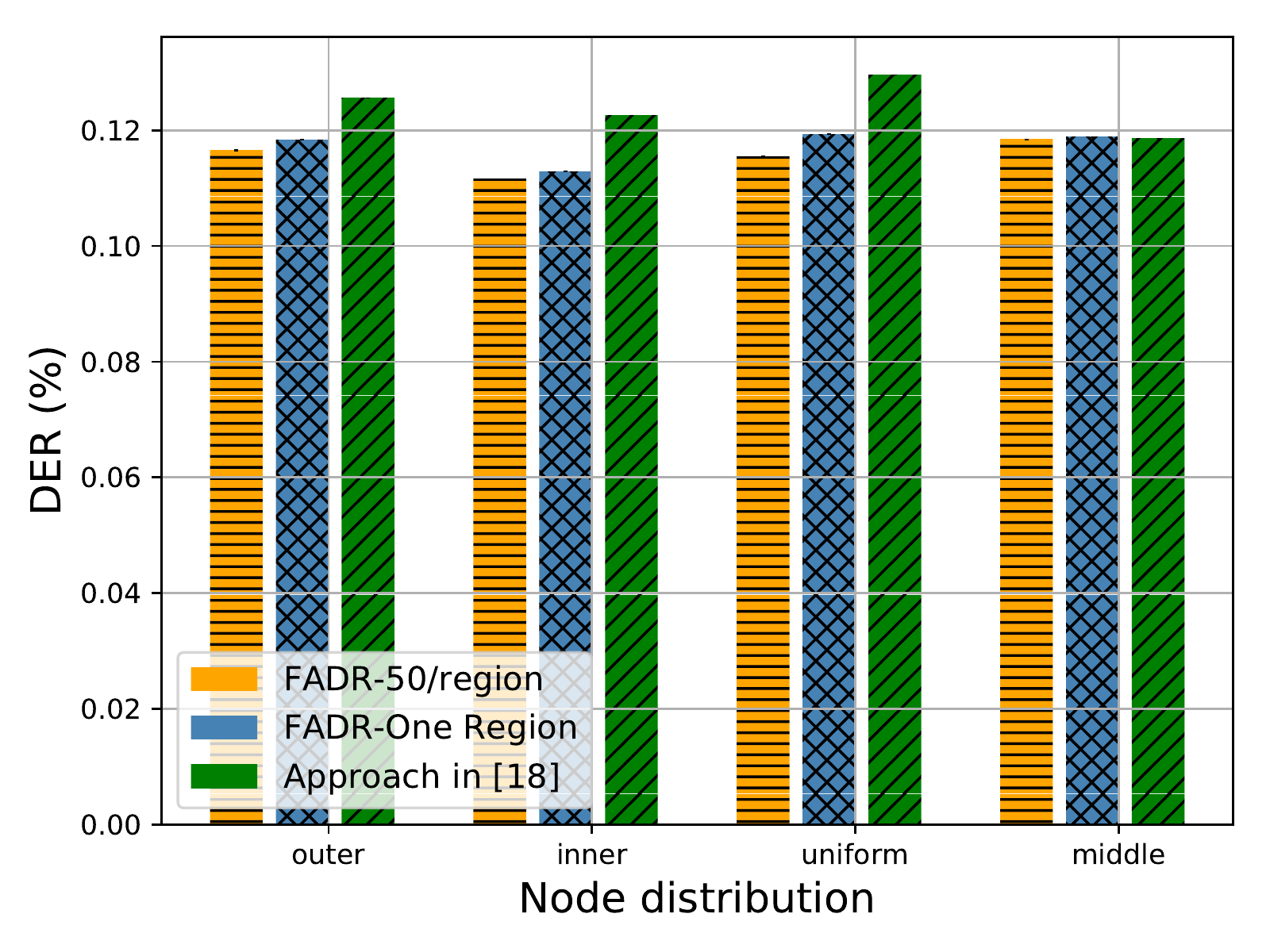}\label{fig:distr_der}
  }
  \subfloat[Fairness Index]{
      \includegraphics[width=0.6\columnwidth]{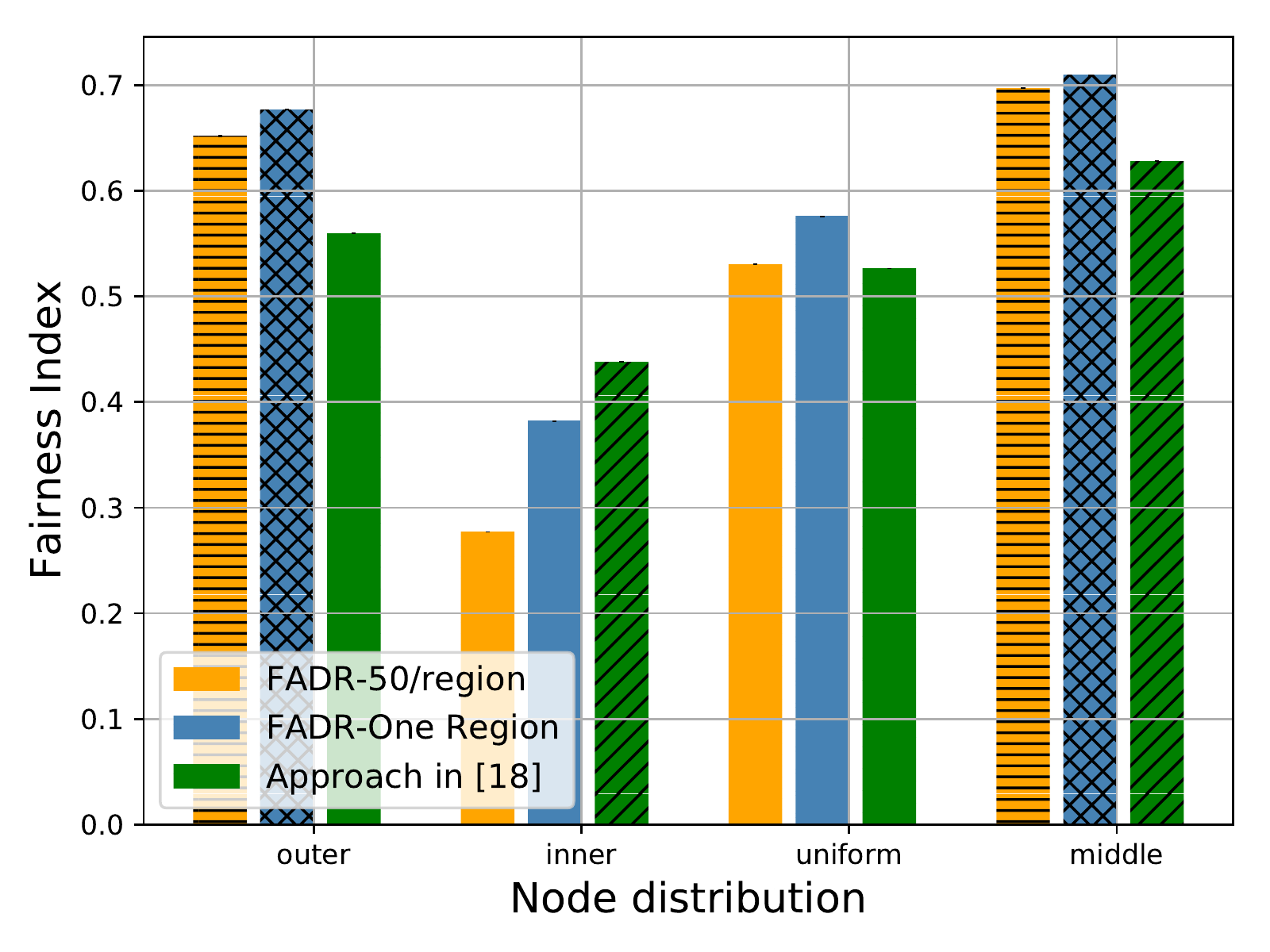}\label{fig:distr_fairness}  
  }
  \subfloat[Energy Consumption]{
      \includegraphics[width=0.6\columnwidth]{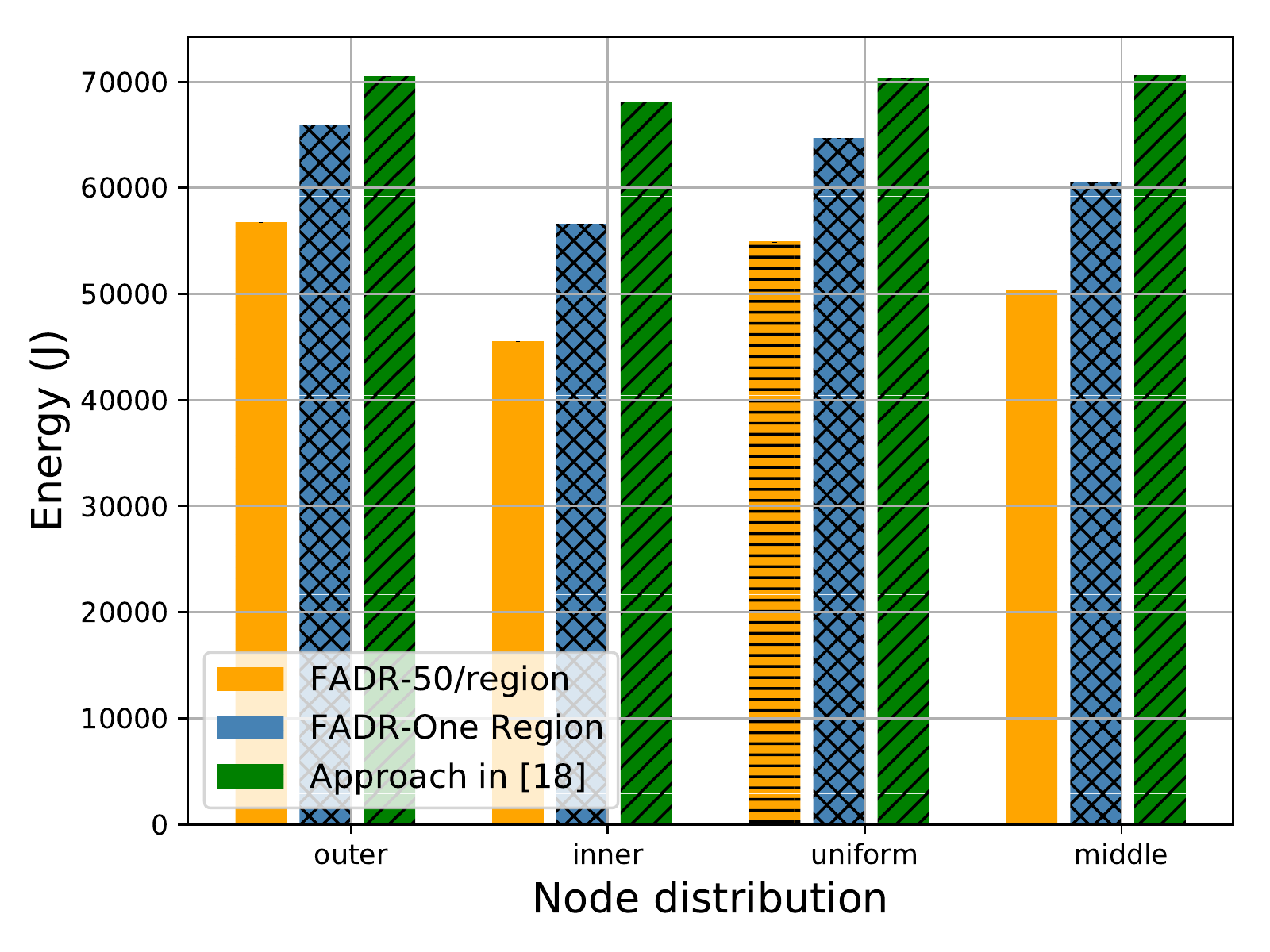}\label{fig:distr_energy}
  }
  \vspace{-0.5em}
  \caption{Node Distribution study}\label{fig:nodediststudy}
  \vspace{-0.5em}
\end{figure*}
   
\subsubsection{Node Distribution Study}\label{nodedistribution}

We used the node distribution implemented in LoRaSim in all aforementioned studies, which randomly distributes the nodes around the gateway. Changing the node distribution will affect the performance of either algorithms. in regard of \cite{reynders2017power}, it changes the location of the nodes with $SF8$, which is the reference for this approach. In regard to FADR, it changes the distribution of collisions between nodes. Therefore, to investigate the impact of different node distributions, the cell is divided into three areas (inner, middle and outer), each 0.33 of the cell radius. Then the distribution of the nodes was adjusted to allocate 66.6\% of nodes to one area and the rest was uniformly distributed in the other two areas. Therefore, the inner distribution, for example, has 66.6\% of nodes in the inner 33\% of the cell radius. Fig.~\ref{fig:nodediststudy} shows the DER (Fig.~\ref{fig:distr_der}), Fairness (Fig.~\ref{fig:distr_fairness}), and Energy consumption (Fig.~\ref{fig:distr_energy}) in the different node distributions. The results shown were collected from a cell with 4000 nodes in each distribution.

Overall, the results validate the observations made so far that the approach in \cite{reynders2017power} achieves higher DER, but higher energy consumption and lower fairness than FADR with the exception of the inner distribution, where FADR achieves lower fairness. Most of the unfairness in Reynders' approach \cite{reynders2017power} comes from the impact of the non-orthogonality of low $SFs$ over high $SFs$ in which \cite{reynders2017power} has higher collisions than FADR. Therefore, FADR from an overall point of view is more suitable for high $SFs$, i.e. edge nodes, compared to \cite{reynders2017power}. However, this comes at the expense of DER in the area close to the gateway, where \cite{reynders2017power} achieves higher DER than FADR.

The inner distribution case is stressful for both approaches because most of the nodes are placed in the high path loss region around the gateway, which affects the remaining 33\% of nodes in the rest of the network. The unfairness stems mostly from the impact of non-orthogonality in which \cite{reynders2017power} has 2.6 times more packets affected by this than FADR. Nevertheless those packets are concentrated in nodes with $SF10-12$ on the outer region of the network. This is because the power boost in nodes with $SF8-9$ is now closer to the gateway, creating a big difference in RSSI larger than the CIR threshold for the nodes in $SF10-12$. Therefore, \cite{reynders2017power} achieves slightly higher DER in nodes with $SF8-9$, but \textit{zero} DER in nodes with $SF10-12$. Whereas, FADR achieves uniformly distributed DER albeit slightly lower over all those nodes. Due to $SF8-9$ being used by more nodes than $SF10-12$, \cite{reynders2017power} achieves slightly higher fairness than FADR. However, if the nodes with $SF7$ are not considered, which have much higher DER than all the remaining $SFs$, FADR achieves 76\% fairness, whereas \cite{reynders2017power} achieves only 64\%.

\subsection{Discussion} \label{discussions}
\subsubsection{Scalability of Fairness} \label{fairnessscability}
From the above studies, it should be noted that increasing the number of nodes, i.e. increasing the number of collisions, has a negative impact on the cell fairness. As LoRaWAN has discrete, limited number of $TPs$ ($2-14dBm$), a cell cannot totally eliminate all collisions using a $TP$ control mechanism. This leads to collisions being not uniformly distributed over distance, but concentrated in certain areas. We see this in the impact of the region of high path loss increase near to the gateway over the rest of the cell. Therefore, increasing the number of collisions magnifies this non-uniformity of collisions, thus, amplifies the unfairness within a cell. Since the transmission rate and the packet length have an impact on the number of collisions as well, these factors affect the fairness as well. While we showed results in this work based on a generated traffic of 80 byte long packets generated once per minute, we found that increasing the transmission rate or packet length with the same number of nodes degrades the fairness.
\subsubsection{Real World Considerations} \label{realconsideration}
The effectiveness of FADR $TP$ control in a real world implementation is affected by the variability of the RSSI, which is not totally stable over time \cite{aref2014free}. Therefore, to avoid RSSI instability, the algorithm is run after a certain number of packets have been collected by the network server to average over RSSI samples. The number of packets that the algorithm should consider before running is under investigation as a future work. Furthermore, it is known that the RSSI values are highly correlated with the propagation model. We used the same log-distance propagation model as in the state-of-art work we compared our approach to. However, we argue the propagation model should not aggressively affect FADR's behavior because FADR does not depend on the RSSI values, but the difference between RSSI values, making FADR more relevant for real world implementations than other approaches that depend on the path-loss estimation. However, as a future work we plan to test FADR's behavior using different propagation models and in real world deployments.

\section{Conclusion} \label{conclusion}
We proposed FADR to achieve a fair data extraction rate in LoRaWAN cells by deploying the fairest data rate ratios that achieve equal collision probability and by controlling transmission power such that it balances the nodes' received power within a safe margin, thus mitigating the capture effect. FADR achieves an almost uniform data extraction rate for all nodes regardless of their distances from the gateway and maintains the nodes' lifetime by not using excessively high transmission power levels. We implemented and compared FADR to other relevant state-of-art work for various network configurations, which showed FADR's advantages.


\section*{Acknowledgment}
This publication has emanated from research conducted with the financial support of Science Foundation Ireland (SFI) and is co-funded under the European Regional Development Fund under Grant Number 13/RC/2077.



\bibliographystyle{IEEEtran}
\bibliography{biblist}
%



\end{document}